\documentclass[final,a4paper,onecolumn]{article}
\usepackage{graphicx,stmaryrd}
\usepackage[usenames]{color}
\usepackage{natbib}
\usepackage{setspace}
\def\nn{\nonumber}
\def\ni{\noindent}
\def\scs{\scriptstyle}
\def\scscs{\scriptscriptstyle}

\title{OpenDDA: A Novel High-Performance Computational Framework for the Discrete Dipole Approximation}

\author{James Mc Donald\footnotemark[1] \and Aaron Golden\footnotemark[1] \and S. Gerard Jennings\footnotemark[2]}

\footnotetext[1]{Computational Astrophysics Laboratory, I.T. Department, National University of Ireland, Galway}
\footnotetext[2]{School of Physics \& Environmental Change Institute, National University of Ireland, Galway}

\begin{document}

\maketitle

\newpage
\begin{center}
\vspace{4em}
PROPOSED RUNNING HEAD: OPENDDA: HIGH-PERFORMANCE DDA\\\vspace{6em}
James Mc Donald\footnotemark[1]\footnotetext[1]{Corresponding author: James Mc Donald (james@it.nuigalway.ie)}\\
Computational Astrophysics Laboratory, Department of Information Technology,\\National University of Ireland, Galway\\
Email: james@it.nuigalway.ie Phone: 353 91 495632 Fax: 353 91 494584\\\vspace{3em}
Aaron Golden\\
Computational Astrophysics Laboratory, Department of Information Technology,\\National University of Ireland, Galway\\
Email: agolden@it.nuigalway.ie Phone: 353 91 493549 Fax: 353 91 494501\\\vspace{3em}
S. Gerard Jennings
\\School of Physics \& Environmental Change Institute,\\National University of Ireland, Galway\\
Email: gerard.jennings@nuigalway.ie Phone: 353 91 492704 Fax: 353 91 495515
\end{center}

\newpage

\begin{abstract}
This work presents a highly-optimised computational framework for the Discrete Dipole Approximation, a numerical method for calculating the optical properties associated with a target of arbitrary geometry that is widely used in atmospheric, astrophysical and industrial simulations. Core optimisations include the bit-fielding of integer data and iterative methods that complement a new Discrete Fourier Transform (DFT) kernel, which efficiently calculates the matrix-vector products required by these iterative solution schemes. The new kernel performs the requisite 3D DFTs as ensembles of 1D transforms, and by doing so, is able to reduce the number of constituent 1D transforms by $60\%$ and the memory by over $80\%$. The optimisations also facilitate the use of parallel techniques to further enhance the performance. Complete OpenMP-based shared-memory and MPI-based distributed-memory implementations have been created to take full advantage of the various architectures. Several benchmarks of the new framework indicate extremely favourable performance and scalability.

\end{abstract}

\ni

\pagestyle{myheadings}
\thispagestyle{plain}

\newpage
\section{Introduction}\label{intro}

\par
The Discrete Dipole Approximation (DDA) is an intuitive and extremely flexible numerical method for calculating the optical properties associated with a target of arbitrary geometry, whose dimensions are comparable to or moderately larger than the incident wavelength. The basic concept was first introduced in 1964 to study the optical properties of molecular aggregates \citep{devoe1964,devoe1965}. Subsequently, prompted by the polarisation of light by interstellar dust and the desire to calculate the approximate extinction, absorption and scattering cross sections for wavelength-sized dielectric grains of arbitrary shape, a comparable scheme was formulated \citep{purcell1973}. Since then, the DDA has undergone major development. A recent erudite review describes this development in detail \citep{yurkin2007}.

\par
There are currently two main publicly available DDA implementations, DDSCAT\footnotemark[3]\footnotetext[3]{http://www.astro.princeton.edu/$\sim$draine} and ADDA\footnotemark[4]\footnotetext[4]{http://www.science.uva.nl/research/scs/Software/adda/index.html}, the development of which has provided seminal contributions to the advancement of the DDA. DDSCAT \citep{draine2004} is written in FORTRAN 77 and is, at present, the most widely used DDA implementation. It has been developed by Bruce Draine from the Princeton University Observatory and Piotr Flatau from the Scripps Institution of Oceanography, University of California, San Diego. Its development provided several advancements including, among others, an updated expression for the dipole polarisability that incorporated a radiative reaction correction \citep{draine1988}, the application of DFT techniques to the iterative core \citep{goodman1991}, and the introduction of the Lattice Dispersion Relations (LDR) for the prescription of the dipole polarisability \citep{draine1993}. A review paper by the authors concisely summarises these contributions \citep{draine1994}. ADDA \citep{yurkin2007b} is an extremely advanced C code that has been developed (over a period of more than 10 years) by Maxim Yurkin and Alfons Hoekstra at the University of Amsterdam. It was the first parallel implementation of DDA and from its initial incarnation has been able to perform single DDA computations across clusters of autonomous computers \citep{hoekstra1994,hoekstra1995,hoekstra1998}. This enables the code to overcome the restrictions on the number of dipoles that can be used due to the memory limitations of a single computer. The code also incorporates several integrated iterative solvers, several prescriptions for prescribing the dipole polarisabilities, and a host of other functionality \citep{yurkin2008}. Other DDA implementations, which are not publicly available, see \citep{yurkin2007b} for further details, include SIRRI \citep{rahola1996,lumme1998}; a FORTRAN 90 code that has been developed by Simulintu Ltd in Finland, and ZDD \citep{zubko2003}; a C++ code that has been developed by Evgenij Zubko and Yuriy Shkuratov from the Institute of Astronomy, Kharkov National University in the Ukraine. For interested parties, a recent paper performs an in-depth comparison of these four codes \citep{penttila2007}.

\par
Put succinctly, the DDA is an extremely flexible and powerful numerical scheme which replaces a continuum scatterer by an ensemble of $N_{d}$ dipoles situated on a cubic lattice. Each of the $N_{d}$ dipoles is driven by the incident electric field and by contributions from the $N_{d}-1$ other dipoles. Eq.~\ref{ddaequations} outlines the kernel of this scheme which is a very large and fully populated set of $3N_{d}$ complex linear equations.

\vspace{-1em}
\begin{eqnarray}
&&\frac{{{\rm k}}^{2}}{\epsilon_{0}}\sum_{j=0}^{N_{d}-1}{\bf A}_{ij}{\bf P}_{j}={\bf E}_{inc,i}\:\:i=0,\ldots,N_{d}-1 \label{ddaequations} \\
&&{\bf A}_{ii}={\alpha_{i}}^{-1} \nn \\
&&{\bf A}_{ij}=\frac{e^{\imath {\rm k}r_{ij}}}{4\pi {r_{ij}}^{3}}
\left\{
\left[\frac{\imath {\rm k}r_{ij}-1}{{{\rm k}}^{2}{r_{ij}}^{2}}\right]
\left[3{\bf r}_{ij}\varotimes{\bf r}_{ij}-{r_{ij}}^{2}{\bf I}\right]+
\left[{\bf r}_{ij}\varotimes{\bf r}_{ij}-{r_{ij}}^{2}{\bf I}\right]
\right\} \nn
\end{eqnarray}

\par
Eq.~\ref{ddaequations} must be solved for the unknown polarisations, ${\bf P}_{j}$, from which the scattering properties can be calculated. Here, ${\rm k}$ is the wavenumber, $\epsilon_{0}$ is the permittivity of free space, $\alpha_{i}$ is the polarisability of dipole $i$, ${\bf E}_{inc,i}$ is the incident electric field at dipole $i$, ${\bf I}$ is the identity matrix, ${\bf r}_{ij}={\bf r}_{i}-{\bf r}_{j}$, $r_{ij}=|{\bf r}_{ij}|$ and ${\bf r}_{ij}\varotimes{\bf r}_{ij}$ is the dyadic product, which is a special case of the Kronecker product for vectors of equal order.

\vspace{-0.5em}
\begin{eqnarray}
{\bf r}_{ij}\varotimes{\bf r}_{ij}=\pmatrix{
{r_{x}}^{2}&r_{x}r_{y}&r_{x}r_{z}\cr
r_{y}r_{x}&{r_{y}}^{2}&r_{y}r_{z}\cr
r_{z}r_{x}&r_{z}r_{y}&{r_{z}}^{2}} \label{dyadicprod}
\end{eqnarray}

\par
For systems based on three-dimensional lattices, the computational requirements are extremely sensitive to the size of the lattice used. Even a small increase in the dimensions of the lattice significantly increases the computational overheads. As a result, the true potential of the host algorithm is often encumbered by the severe and often prohibitive burdens imposed by the computational scale required to produce meaningful results. In these cases, it is crucial that every possible algorithmic or computational enhancement be exploited in order to augment their applicability. OpenDDA is a novel computational framework for the DDA which has been custom-built from scratch in the C language. The new framework incorporates a variety of algorithmic and computational optimisations, enabling researchers to take full advantage of the different architectures that exist in the current generation of computational resources and beyond. Apart from the core optimisations, the new framework also incorporates an extremely intuitive and user-friendly interface for specifying the simulation parameters, including a sturdy and forgiving parser and a straightforward human-readable input control file.

\par
The paper is organised as follows. Section \ref{bitfieldtarocccompdata} discusses the application of bit-fielding to streamline the storage of the target occupation and composition data. Section \ref{kernelopt} deals with the design and development of a new highly-optimised DFT kernel to calculate the matrix-vector products within the iterative solution schemes. Section \ref{itersch} discusses the custom-built iterative schemes that have been included to support the new DFT kernel. Section \ref{parallelimp} describes the extension of this formulation to both shared-memory and distributed-memory parallel environments. Section \ref{benchmark} provides various various serial and parallel benchmarks to corroborate the assertions and illustrate the implications of the new optimisations. Finally, Section \ref{conclusions} presents concluding remarks.

\section{Bit-fielding of target occupation and composition data}\label{bitfieldtarocccompdata}

\par
In many applications, the range provided by the standard integer data types far exceeds the maximum range associated with the task at hand, and as a result, for large datasets, most of the bits are superfluous, constituting a serious waste of resources. Bit-fielding is a way to overcome this issue by storing data within the bits rather than the integers themselves. The entries can be set \& tested by taking the appropriately shifted bit-mask and applying the relevant bitwise operator, LOGICAL OR for setting \& LOGICAL AND for testing. The simplicity and speed of bitwise operations makes bit-fielding an extremely efficient method of storing non-negative integer based data.

\par
The DDA's theoretical strength, which is also its computational weakness due to imposed computational overheads, is its volumetric nature. Its ability to treat arbitrarily shaped particles with complex internal structures derives from the fact that each constituent site in the computational domain has separate integer flags describing its properties. Rather than storing the spatial extent of the target as the Cartesian coordinates of each lattice site in the computational domain and a Boolean occupation flag to indicate whether or not the site is occupied, requiring four integers, OpenDDA simply stores a single bit-fielded occupation flag per lattice site. The Cartesian coordinates need not be stored explicitly as, if required, are readily available as the loop control variables or via direct extraction from the array index itself. The optimisation of this integer data storage reduces the memory required for the spatial description of the target by a factor of 128.

\par
The DDA also requires the specification of the material properties of the target on a dipole-by-dipole basis. This is done by associating each complex refractive index value with a distinct integer flag, and for each dipole, storing three integers to define the material properties in the ${\bf\hat x}$, ${\bf\hat y}$ \& ${\bf\hat z}$ directions. To compress this composition data, OpenDDA employs an extension of the bit-fielding theory to non-Boolean data sets. The new framework automatically calculates the requisite bit-mask, i.e., for four disparate materials, $0\to3$, two bits are all that is required, and the bit-mask is $3_{dec}=11_{bin}$. For this example, the memory required for the integer data associated with the compositional description of the target is reduced by a factor of 16.

\par
To put the optimisations in perspective, for $K=J=P=500,\:N=K\times J\times P=125\times10^{6}$, where $K$, $J$ \& $P$ are the ${\bf\hat x}$, ${\bf\hat y}$ \& ${\bf\hat z}$ dimensions of the computational domain, respectively, with four materials, the memory requirements for the integer data associated with the spatial and compositional description of the target are reduced from 3.26${\rm GB}$ to just over 0.1${\rm GB}$. Furthermore, the computational overhead incurred by the application of this bit-fielding optimisation within OpenDDA is negligible. Firstly, OpenDDA does not require the actual Cartesian coordinates within the iterative core. They are only needed at the initialisation stage for the calculation of the incident electric field at each of the lattice sites. Secondly, due to the nature of the iterative methods and the fact that the iterative solution vectors only include entries corresponding to occupied sites (discussed in \S\ref{itersch}), the Boolean occupation flags are only tested at the initialisation stage, during the calculation of the dipole polarisabilities, and once at the start and end of the DFT matrix-vector product routine. Although the integer data constitutes only a fraction of the total memory footprint, with most deriving from the requisite complex vectors, these optimisations are not without merit and fundamentally underpin the efficiency of the current distributed-memory implementation of OpenDDA. To facilitate the elimination of zero storage within all iterative solution vectors, each autonomous node requires local knowledge of the entire spatial and compositional specification of the target and thus, unlike the complex vectors within the iterative core, the target specification is not distributed. Bit-fielding allows the full description of the target to be known globally without a prohibitive overhead. With each node having typically a few gigabytes to work with, the original memory requirements associated with the description of the target would have been comparable to the maximum possible allocation per node, negating the possibility of a viable distributed implementation.

\section{Kernel optimisation}\label{kernelopt}

\par
Due to the sheer size of the system produced by Eq.~\ref{ddaequations}, iterative $O(h[g]^{2})$ methods, where $g=3N_{d}$ and $h$ is the number of iterations required to reach convergence, are employed, which are much faster than direct $O(g^{3})$ methods for large $N_{d}$ and $h\ll N_{d}$. The cardinal computational hurdle and key to the acceleration of these iterative methods, such as the conjugate gradient method \citep{barrett1994,golub1996}, are the matrix-vector products required by these schemes. A major breakthrough in this respect was the realisation that by embedding the arbitrarily shaped target in a $N=K\times J\times P$ computational domain and neglecting the diagonal terms, the matrix-vector products can be reformulated as convolutions \citep{goodman1991}. Once the convolutions have been performed, it is then a trivial matter to include the contributions from the previously neglected terms.

\par
The solution method is predicated upon Toeplitz and circulant matrix structure \citep{davis1994,golub1996}. A Toeplitz matrix of order $[n]$, denoted ${}^{{\scscs[n]}}{\rm T}$, is simply a matrix with constant diagonals, whereas a circulant matrix, ${}^{{\scscs[n]}}{\rm C}$, is a special case of Toeplitz structure, where the last element of each row is the same as first element of the next row. Any Toeplitz matrix, ${}^{{\scscs[n]}}{\rm T}$, can be converted to its circulant counterpart, ${}^{{\scscs[n^{\prime}]}}{\rm C}$, where $n^{\prime}=2n-1$, by taking the first column, appending the entries of the first row, excluding the initial entry, in reverse order, and completing the diagonals. The key is that any circulant matrix is diagonalised by the corresponding Fourier matrix, ${}^{{\scscs[n^{\prime}]}}{\rm F}$ \citep{vanloan1992}, where ${}^{{\scscs[n^{\prime}]}}{\rm c}$ is the first column of ${}^{{\scscs[n^{\prime}]}}{\rm C}$.

\vspace{-1em}
\begin{eqnarray}
&&{}^{{\scscs[n^{\prime}]}}{\rm F}=\frac{1}{\sqrt{n^{\prime}}}\omega^{pq},\:p,q=0,\ldots,n^{\prime}-1,\:\omega=e^{-\frac{2\pi\imath}{n^{\prime}}} \nn \\
&&{}^{{\scscs[n^{\prime}]}}{\rm F}{}^{{\scscs[n^{\prime}]}}{\rm C}{}^{{\scscs[n^{\prime}]}}{\rm F}^{\rm H}=diag\left(\sqrt{n^{\prime}}\left[{}^{{\scscs[n^{\prime}]}}{\rm F}\right]{}^{{\scscs[n^{\prime}]}}{\rm c}\right)=diag\left(\lambda_{C}\right) \label{circdiagl0}
\end{eqnarray}

\par
Using the fact that the Fourier matrix is unitary, ${}^{{\scscs[n^{\prime}]}}{\rm F}^{-1}={}^{{\scscs[n^{\prime}]}}{\rm F}^{\rm H}$, the multiplication of an arbitrary vector with a circulant matrix can be performed as an element-wise multiplication in the Fourier Domain (FD) via the Convolution Theorem, $f\rightleftharpoons{\rm F},\:g\rightleftharpoons{\rm G},\:f*g\rightleftharpoons{\rm F}.{\rm G}$. The symbol `$*$' denotes convolution, `$\rightleftharpoons$' denotes a Fourier transform pair and the symbol `$.$' implies element-wise multiplication.

\vspace{-1em}
\begin{eqnarray}
{}^{{\scscs[n^{\prime}]}}{\rm y}={}^{{\scscs[n^{\prime}]}}{\rm C}{}^{{\scscs[n^{\prime}]}}{\rm x}={}^{{\scscs[n^{\prime}]}}{\rm F}^{-1}diag\left(\lambda_{C}\right){}^{{\scscs[n^{\prime}]}}{\rm F}{}^{{\scscs[n^{\prime}]}}{\rm x}={}^{{\scscs[n^{\prime}]}}{\rm F}^{-1}[\lambda_{C}].[{}^{{\scscs[n^{\prime}]}}{\rm F}{}^{{\scscs[n^{\prime}]}}{\rm x}] \label{circdiagl1}
\end{eqnarray}

\par
Not only does this reduce the computational complexity of the matrix-vector product from $O(n^{2})$ to $O(n^{\prime}\ln n^{\prime})$, but since the circulant eigenvalues are a scalar multiple of the DFT of the first column of the circulant matrix, see Eq.~\ref{circdiagl0}, the memory is reduced from $O(n^{2})$ to $O(n^{\prime})$. To expedite the calculation of a Toeplitz matrix-vector product, ${}^{{\scscs[n]}}{\rm y}={}^{{\scscs[n]}}{\rm T}{}^{{\scscs[n]}}{\rm x}$, the matrix must undergo Toeplitz-to-circulant (T-to-c) conversion, ${}^{{\scscs[n]}}{\rm T}\to{}^{{\scscs[n^{\prime}]}}{\rm C}$, and the vector zero-padded to the dimension of the circulant matrix by appending $n-1$ zeroes. Eq.~\ref{circdiagl1} can then be applied. The first $n$ elements of the resultant are the required answer for the original Toeplitz matrix-vector product, with the remaining $n-1$ entries being redundant. Please note that this methodology and the following discussion is only applicable to systems exhibiting Toeplitz structure. Non-Toeplitz structure negates the transformation to circulant structure and hence, the use of DFTs to calculate the matrix-vector products efficiently. In terms of the specific geometry of the target in question, the method is indifferent. For sparse systems, from targets with a low volume filling fraction (VFF), such as dendritic or highly porous structures, the only difference is the amount of memory required by the iterative solution vectors. The solution vectors within the iterative core do not store entries corresponding to vacant lattice sites, a fact that will be discussed in \S\ref{itersch}. However, the DFT kernel that efficiently calculates the matrix-vector products required by the iterative solvers must account for the entire rectangular computational domain, including the vacant sites, and thus performs the same amount of calculations, irrespective of the VFF.

\par
For the DDA, by embedding the arbitrary target in a rectangular lattice and neglecting the diagonal terms, the requisite matrix-vector products take on the form of a complex-symmetric level-three tensor-Toeplitz matrix of order $[P][J][K]$, denoted ${}^{{\scscs[P][J][K]}}{\rm T}$, in which, the constituent tensor elements are complex-symmetric but not necessarily Toeplitz, multiplied by an arbitrary vector of length $N$, denoted ${}^{{\scscs[N]}}{\rm x}$, whose constituent elements are Cartesian vectors. Computationally, ${}^{{\scscs[P][J][K]}}{\rm T}$ can be considered as an assemblage of nine complex-symmetric level-three single-element Toeplitz matrices, and since the tensor elements are symmetric, only six of the nine tensor components are independent and need to be considered explicitly, $xx,xy,xz,yy,yz,zz\equiv$${}^{{\scscs[P][J][K]}}{\rm T0}$, ${}^{{\scscs[P][J][K]}}{\rm T1},\ldots,{}^{{\scscs[P][J][K]}}{\rm T5}$. Analogously, the arbitrary vector can be split into three separate $x,y,z$ component vectors, ${}^{{\scscs[N]}}{\rm x0}$, ${}^{{\scscs[N]}}{\rm x1}$, ${}^{{\scscs[N]}}{\rm x2}$. Consequently, this means that the components can be treated separately and the full problem is reduced to an ensemble of smaller single-element problems whose contributions will ultimately be manifested through FD products.

\par
At this point, before attempting to decipher the ensuing optimisations of this DFT technique, for readers who are not familiar with the specific algorithmic details associated with the application of DFTs to the solution of such matrix-vector systems, it would be prudent to read Appendix \ref{origdftalgor}. This concisely summarises the conventional application of DFTs to these complex-symmetric level-three tensor-Toeplitz systems in the context of the DDA, as prescribed by \cite{goodman1991}.

\par
The complete standard algorithm, which encapsulates the conventional implementation, can be summarised as follows:

\begin{enumerate}
\item T-to-c conversion and 3D DFT of the six independent tensor components
\vspace{-1em}
\item Zero pad and 3D DFT of the three zero-padded vector components
\vspace{-1em}
\item FD element-wise multiplication of the tensor and vector components
\vspace{-1em}
\item 3D iDFT and normalisation of the components of the resultant
\vspace{-1em}
\item Addition of the previously ignored diagonal contributions
\end{enumerate}

\par
\begin{figure}[p]
\begin{center}
\includegraphics[scale=0.75]{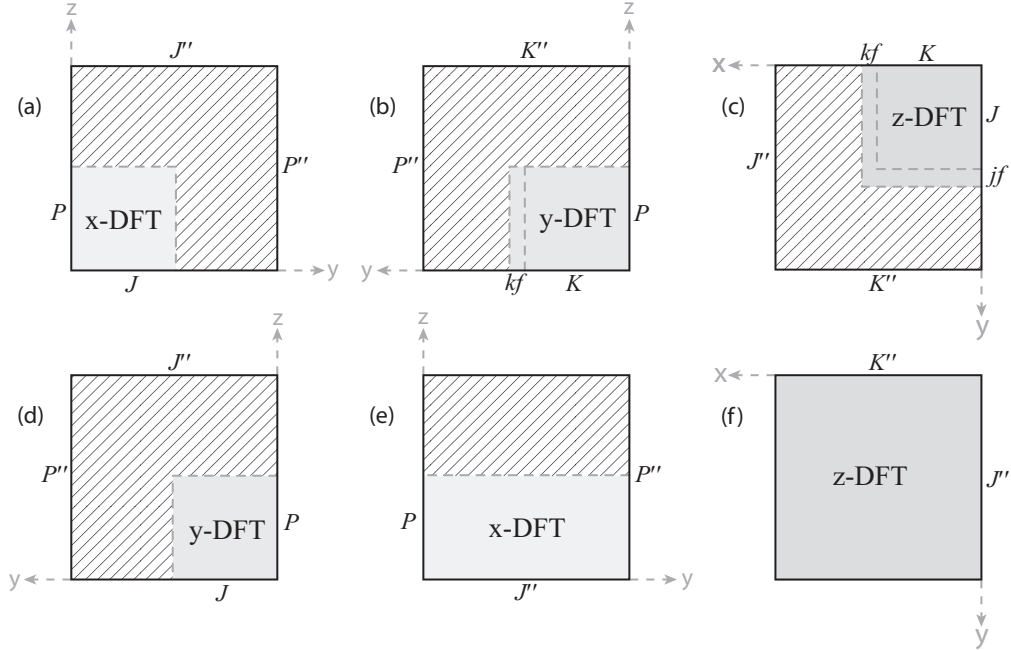}
\vspace{-1em}
\caption{\label{fig1} Outlines the optimisation of the requisite 3D DFTs via the implicit T-to-c conversion and simultaneous DFT of the tensor-component arrays and, for the vector components, by only performing transforms that are non-zero or that will contribute to the final answer (a) Tensor components (${\bf {\hat x}}$ direction) (b) Tensor components (${\bf {\hat y}}$ direction) (c) Tensor components (${\bf {\hat z}}$ direction) (d) Vector components (${\bf {\hat y}}$ direction) (e) Vector components (${\bf {\hat x}}$ direction) (f) Vector components (${\bf {\hat z}}$ direction).}
\end{center}
\end{figure}

Due to the T-to-c conversions that are required for DFT applicability, the six tensor-components arrays contain a high degree of symmetry. Since a $K^{\prime}\times J^{\prime}\times P^{\prime}$ 3D DFT can be carried out as an ensemble of 1D DFTs, $J^{\prime}\times P^{\prime}$ in the ${\bf {\hat x}}$, $K^{\prime}\times P^{\prime}$ in the ${\bf {\hat y}}$, and $K^{\prime}\times J^{\prime}$ in the ${\bf {\hat z}}$ direction, a new algorithm which exploits these symmetries can be constructed. Rather than expanding the six tensor-component arrays from a $K\times J\times P$ Toeplitz structure to a $K^{\prime}\times J^{\prime}\times P^{\prime}$ circulant structure and applying 3D DFTs, 
as delineated in Fig.~\ref{fig6}, the T-to-c conversion and subsequent DFT can be amalgamated into a single computationally efficient operation. Each of the lines in the ${\bf {\hat x}}$ direction, ${}^{{\scscs[K]}}{\rm t0{\scs 3D}}_{j,k},j=0,\ldots,J-1,k=0,\ldots,P-1$, in the ${\bf {\hat y}}$ direction, ${}^{{\scscs[J]}}{\rm t0{\scs 3D}}_{i,k},i=0,\ldots,K-1,k=0,\ldots,P-1$, and in the ${\bf {\hat z}}$ direction, ${}^{{\scscs[P]}}{\rm t0{\scs 3D}}_{i,j},i=0,\ldots,K-1,j=0,\ldots,J-1$, in section `$1$' in Fig.~\ref{fig6}, can be taken in turn and copied into an external 1D array of length $K^{\prime}=2K-1$, $J^{\prime}=2J-1$, and $P^{\prime}=2P-1$, respectively. Each can be treated as the first column of a level-one Toeplitz matrix and the T-to-c conversion performed. The DFTs of these external vectors can be computed, and subsequently, the first $K$, $J$ and $P$ elements of the resultants can be copied back into the 3D tensor-component array, overwriting the original data. This significantly reduces the number of 1D DFTs that have to be performed. Taking into account the zero-padding to ensure that $K^{\prime\prime}=K^{\prime}+kf=2K-1+kf$, $J^{\prime\prime}=J^{\prime}+jf=2J-1+jf$, and $P^{\prime\prime}=P^{\prime}+pf=2P-1+pf$ permits the use of fast DFT algorithms, the number of transforms is reduced by approximately $75\%$ in total:

\doublespace
\begin{tabbing}
\hspace{0.5em}\=${\bf {\hat x}}$ {\bf direction, see Fig.~\ref{fig1}(a):}\hspace{0.5em}\=$6\left[J^{\prime\prime}\times P^{\prime\prime}\right]$\hspace{0.5em}\=$\sim\:$\fcolorbox{Gray}{White}{${\bf 24}\left[J\times P\right]$}\hspace{0.5em}\=$\rightarrow$\hspace{0.5em}\=\fcolorbox{Gray}{White}{${\bf 6}\left[J\times P\right]$} \\
\>${\bf {\hat y}}$ {\bf direction, see Fig.~\ref{fig1}(b):}\hspace{0.5em}\>$6\left[K^{\prime\prime}\times P^{\prime\prime}\right]$\>$\sim\:$\fcolorbox{Gray}{White}{${\bf 24}\left[K\times P\right]$}\>$\rightarrow$\>$6\left[(K+kf)\times P\right]$$\sim\:$\fcolorbox{Gray}{White}{${\bf 6}\left[K\times P\right]$} \\
\>${\bf {\hat z}}$ {\bf direction, see Fig.~\ref{fig1}(c):}\hspace{0.5em}\>$6\left[K^{\prime\prime}\times J^{\prime\prime}\right]$\>$\sim\:$\fcolorbox{Gray}{White}{${\bf 24}\left[K\times J\right]$}\>$\rightarrow$\>$6\left[(K+kf)\times(J+jf)\right]$$\sim\:$\fcolorbox{Gray}{White}{${\bf 6}\left[K\times J\right]$}
\end{tabbing}
\singlespace

\par
Furthermore, since the full T-to-c conversion, as delineated in Fig.~\ref{fig6}, is never explicitly performed, only section `1' in Fig.~\ref{fig6} need be stored, reducing the complex number storage required for the six tensor-component arrays by approximately $88\%$:

\begin{tabbing}
\hspace{1em}\=$6\left[K^{\prime\prime}\times J^{\prime\prime}\times P^{\prime\prime}\right]\sim\:$\fcolorbox{Gray}{White}{${\bf 48}\left[K\times J\times P\right]$}$\:\rightarrow6\left[(K+kf)\times(J+jf)\times(P+pf)\right]\sim\:$\fcolorbox{Gray}{White}{${\bf 6}\left[K\times J\times P\right]$}
\end{tabbing}

\par
For the three zero-padded vector-component arrays, the realisation that $\approx\frac{7}{8}$ of each is by very definition, identically zero, can be used to expedite the calculations. This was originally proposed by \cite{hoekstra1998} and has been exploited within ADDA \citep{yurkin2008}. Within OpenDDA, for computational efficiency, the DFTs of the three vector components, the subsequent element-wise FD multiplication of the tensor and vector components, and the iDFTs of the resultant, are amalgamated into a conjoint operation. Unlike the tensor component DFTs, which are performed in the ${\bf {\hat x}}$, ${\bf {\hat y}}$ \& ${\bf {\hat z}}$ directions, the DFTs of the vector components are performed in the ${\bf {\hat y}}$, ${\bf {\hat x}}$ \& ${\bf {\hat z}}$ directions, and the iDFTs are performed in the ${\bf {\hat z}}$, ${\bf {\hat x}}$ \& ${\bf {\hat y}}$ directions. The reason for this is twofold. Firstly, it allows the use of a $xz$-scratch plane for the DFTs in the ${\bf {\hat x}}$ \& ${\bf {\hat z}}$ directions, the FD multiplication, and the iDFTs in the ${\bf {\hat z}}$ \& ${\bf {\hat x}}$ directions, which provides significant memory reductions, as only $\approx\frac{1}{4}$ of each of the three full $K^{\prime\prime}\times J^{\prime\prime}\times P^{\prime\prime}$ arrays need be stored. Secondly, it supports the distributed FD multiplication in the MPI version of the algorithm (discussed in \S\ref{parallelimp}). As a result of this optimisation, the complex number storage for the three vector components, ignoring the storage required for the three $xz$-scratch planes, is reduced by approximately $75\%$:

\begin{tabbing}
\hspace{5em}\=$3\left[K^{\prime\prime}\times J^{\prime\prime}\times P^{\prime\prime}\right]\sim\:$\fcolorbox{Gray}{White}{${\bf 24}\left[K\times J\times P\right]$}$\:\rightarrow\:$ $3\left[K\times J^{\prime\prime}\times P\right]\sim\:$\fcolorbox{Gray}{White}{${\bf 6}\left[K\times J\times P\right]$}
\end{tabbing}

\par
Each of the three vector components is embedded into a zero-padded $\left[K\times J^{\prime\prime}\times P\right]$ array, $K$ in the ${\bf {\hat x}}$ direction, $J^{\prime\prime}$ in the ${\bf {\hat y}}$ direction, and $P$ in the ${\bf {\hat z}}$ direction. Once embedded, the DFT is performed in the ${\bf {\hat y}}$ direction. Then, each of the $\left[K\times P\right]$ $xz$-planes is, in turn, copied into a $\left[K^{\prime\prime}\times P^{\prime\prime}\right]$ external zero-padded scratch $xz$-plane. Subsequently, the DFTs in the ${\bf {\hat x}}$ \& ${\bf {\hat z}}$ directions, the element-wise FD multiplication of the tensor \& vector components, and the iDFTs in the ${\bf {\hat z}}$ \& ${\bf {\hat x}}$ directions are performed `on-the-fly' for this $xz$-plane. The initial $\left[K\times P\right]$ section is then copied back into the 3D vector component array, overwriting the original data. Finally, once this has been done for each of the $J^{\prime\prime}$ $xz$-planes, the iDFT in the ${\bf {\hat y}}$ direction is performed.

\par
The reason that this is all possible is due to the fact that only non-zero DFTs, and only iDFTs that will contribute to the final answer, need to be performed. As far as the DFTs are concerned, in the ${\bf {\hat y}}$ direction, only the lines ${}^{{\scscs[J^{\prime}]}}{\rm x0{\scs 3D}}_{i,k},i=0,\ldots,K-1,k=0,\ldots,P-1$, see Fig.~\ref{fig1} (d), contain any non-zero values and need be considered. Subsequently, in the ${\bf {\hat x}}$ direction, only the lines ${}^{{\scscs[K^{\prime}]}}{\rm x0{\scs 3D}}_{j,k},j=0,\ldots,J^{\prime}-1,k=0,\ldots,P-1$, see Fig.~\ref{fig1} (e), need be transformed, and in the ${\bf {\hat z}}$ direction, all the lines ${}^{{\scscs[P^{\prime}]}}{\rm x0{\scs 3D}}_{i,j},i=0,\ldots,K^{\prime}-1,j=0,\ldots,J^{\prime}-1$, see Fig.~\ref{fig1} (f), contain non-zero values and must be included. Taking into account the zero-padding to ensure that $K^{\prime\prime}=K^{\prime}+kf=2K-1+kf$, $J^{\prime\prime}=J^{\prime}+jf=2J-1+jf$, and $P^{\prime\prime}=P^{\prime}+pf=2P-1+pf$ permits the use of fast DFT algorithms, the number of transforms is reduced by approximately $42\%$ in total:

\doublespace
\begin{tabbing}
\hspace{3em}\=${\bf {\hat y}}$ {\bf direction, see Fig.~\ref{fig6}(d):}\hspace{0.5em}$3\left[K^{\prime\prime}\times P^{\prime\prime}\right]$\hspace{0.5em}\=$\sim\:$\fcolorbox{Gray}{White}{${\bf 12}\left[K\times P\right]$}\hspace{0.5em}\=$\:\rightarrow\:$\hspace{0.5em}\=\fcolorbox{Gray}{White}{$3\left[K\times P\right]$} \\
\>${\bf {\hat x}}$ {\bf direction, see Fig.~\ref{fig6}(e):}\hspace{0.5em}$3\left[J^{\prime\prime}\times P^{\prime\prime}\right]$\>$\sim\:$\fcolorbox{Gray}{White}{${\bf 12}\left[J\times P\right]$}\>$\:\rightarrow\:$\>$3\left[J^{\prime\prime}\times P\right]\sim\:$\fcolorbox{Gray}{White}{$6\left[J\times P\right]$} \\
\>${\bf {\hat z}}$ {\bf direction, see Fig.~\ref{fig6}(f):}\hspace{0.5em}`\emph{no reduction possible}'
\end{tabbing}
\singlespace

\par
The FD element-wise multiplications can be carried out just as detailed in Eqs.~\ref{Q0eqn} $\rightarrow$ \ref{Q2eqn} in Appendix \ref{origdftalgor}, with one minor caveat. The six independent tensor-component arrays never underwent the full T-to-c conversion, however, since the DFT preserves the mirror symmetries, when accessing ${\rm Y0}\rightarrow{\rm Y5}$, the pertinent data can be extracted using the criteria:

\vspace{-1em}
\begin{eqnarray}
\begin{array}{ccc}
i=\cases{i,&if $i<K$;\cr K^{\prime}-i,&if $i\geq K$.\cr}
&
j=\cases{j,&if $j<J$;\cr J^{\prime}-j,&if $j\geq J$.\cr}
&
k=\cases{k,&if $k<P$;\cr P^{\prime}-k,&if $k\geq P$.\cr} \cr
i=0,\ldots,K^{\prime}-1
&
j=0,\ldots,J^{\prime}-1
&
k=0,\ldots,P^{\prime}-1
\end{array} \nn
\end{eqnarray}

\par
For the iDFT of the resultant of the FD multiplication, i.e., of ${\rm Q0}$, ${\rm Q1}$ \& ${\rm Q2}$, see Eqs.~\ref{Q0eqn} $\rightarrow$ \ref{Q2eqn} in Appendix \ref{origdftalgor}, only transforms that will contribute to the final requisite answer need to be performed. In the ${\bf {\hat z}}$ direction, all the lines ${}^{{\scscs[P^{\prime}]}}{\rm Q0}_{i,j},i=0,\ldots,K^{\prime}-1,j=0,\ldots,J^{\prime}-1$, see Fig.~\ref{fig1} (f), contribute, through the subsequent ${\bf {\hat x}}$ and ${\bf {\hat y}}$ transforms, to the final answer. Subsequently, in the ${\bf {\hat x}}$ direction, only the lines ${}^{{\scscs[K^{\prime}]}}{\rm Q0}_{j,k},j=0,\ldots,J^{\prime}-1,k=0,\ldots,P-1$, see Fig.~\ref{fig1} (e), contribute, through the subsequent ${\bf {\hat y}}$ transform, to the final answer, and finally, in the ${\bf {\hat y}}$ direction, only the lines ${}^{{\scscs[J^{\prime}]}}{\rm Q0}_{i,k},i=0,\ldots,K-1,k=0,\ldots,P-1$, see Fig.~\ref{fig1} (d), contribute and need to be considered. Taking into account the zero-padding to ensure that $K^{\prime\prime}=K^{\prime}+kf=2K-1+kf$, $J^{\prime\prime}=J^{\prime}+jf=2J-1+jf$, and $P^{\prime\prime}=P^{\prime}+pf=2P-1+pf$ permits the use of fast iDFT algorithms, the number of transforms is reduced, as with the forward DFTs, by approximately $42\%$ in total:

\doublespace
\begin{tabbing}
\hspace{3em}\=${\bf {\hat z}}$ {\bf direction, see Fig.~\ref{fig6}(f):}\hspace{0.5em}\=`\emph{no reduction possible}'\hspace{-4em}\=\hspace{6.5em}\=\hspace{2em}\= \\
\>${\bf {\hat x}}$ {\bf direction, see Fig.~\ref{fig6}(e):}\hspace{0.5em}\>$3\left[J^{\prime\prime}\times P^{\prime\prime}\right]$\>$\sim\:$\fcolorbox{Gray}{White}{$12\left[J\times P\right]$}\>$\:\rightarrow\:$\>$3\left[J^{\prime\prime}\times P\right]\sim\:$\fcolorbox{Gray}{White}{$6\left[J\times P\right]$} \\
\>${\bf {\hat y}}$ {\bf direction, see Fig.~\ref{fig6}(d):}\hspace{0.5em}\>$3\left[K^{\prime\prime}\times P^{\prime\prime}\right]$\>$\sim\:$\fcolorbox{Gray}{White}{$12\left[K\times P\right]$}\>$\:\rightarrow\:$\>\fcolorbox{Gray}{White}{$3\left[K\times P\right]$}
\end{tabbing}
\singlespace

\ni
The approximate complexity improvements for both computation and storage are given in Table.~\ref{table2}. Please note that for all three architectural variants, serial, OpenMP (shared-memory), and MPI (distributed-memory), the DFT functionality required by the iterative core is provided by the advanced complex interface of the extremely efficient, highly-portable and well known FFTW package (version 3) \citep{fftw3}.

\section{Iterative schemes}\label{itersch}

\par
To avoid unnecessary changes in data format to comply with the requisite structure of some external package, which would only serve to subvert any and all attempts to maximise the performance and streamline the structure of the framework, (like similar DDA implementations) OpenDDA incorporates a relatively large selection of iterative schemes that have been specifically custom-built to integrate seamlessly with the new DFT matrix-vector kernel and support the custom-built domain decomposition and parallel memory allocation schemes in the MPI implementation. Although a variety of different iterative solvers have been employed by \cite{draine1994,lumme1994,flatau1997,nebeker1998,rahola1998,fan2006,penttila2007,yurkin2007b}, to date, no definitive ``best'' algorithm has been identified. To permit further investigation and to maximise the flexibility and usefulness of the end product, OpenDDA includes: Conjugate Gradients ({\bf CG}) \citep{hestenes1952,shewchuk1994}, Conjugate Gradients Squared ({\bf CGS}) \citep{sonneveld1989,barrett1994}, BiConjugate Gradients ({\bf BiCG}) \citep{fletcher1975,barrett1994}, BiCG for symmetric systems (${\bf BiCG}_{\bf sym}$) \citep{freund1992a}, Stabilised version of the BiCG ({\bf BiCGSTAB}) \citep{vandervorst1992}, Restarted, stabilised version of the BiCG ({\bf RBiCGSTAB}) \citep{sleijpen1993}, Quasi-minimal residual with coupled two-term recurrences ({\bf QMR}) \citep{freund1991,freund1992,bucker1996}, QMR for symmetric systems (${\bf QMR}_{\bf sym}$) \citep{freund1992a}, Transpose-free QMR ({\bf TFQMR}) \citep{freund1993} and a variant of BiCGSTAB based on multiple Lanczos starting vectors ({\bf ML(n)BiCGSTAB}) \citep{yeung1999}.

\par
For computational efficiency, as in ADDA \citep{yurkin2007b,yurkin2008}, the new framework only stores information corresponding to occupied lattice sites, i.e., the elimination of zero storage within all iterative solution vectors. For the majority of cases, this provides a significant reduction in the memory required to store these vectors, as the target being simulated does not fully utilise the computational domain, generally having fewer occupied sites towards the exterior. As a quick example, consider a spherical target based on an $18\!\times\!18\!\times\!18$ computational grid. Although the grid contains $5832$ lattice sites, for a spherical target, only $3112$ are occupied, and dumping the unoccupied data, which is identically zero, equates to a decrease of $\sim47\%$ in the memory required for the iterative vectors. To facilitate this optimisation, data extension and compression algorithms were created to extend a target vector out to allow the DFT matrix-vector product, and subsequently, to compress the resultant back into the computationally efficient form with just the entries corresponding to occupied sites.

\section{Parallel implementations}\label{parallelimp}

\par
Over and above the serial improvements, due to the fact that the DFT kernel has been reduced from using 3D to 1D transforms, parallel techniques can be used to further enhance the performance. Full OpenMP-based and MPI-based implementations of the new framework have been created to take proper advantage of both shared-memory and distributed-memory architectures.

\par
For the shared-memory implementation, for thread safety, each thread must have its own DFT scratch arrays. Furthermore, since the aforementioned data-compression algorithm at the end of the DFT kernel operates in-place within the result buffer, to prevent pertinent data loss by multiple processes working within the same array, it has a maximum scalability of three, i.e., one thread per vector component.

\par
\begin{figure}[p]
\begin{center}
\includegraphics[scale=0.5]{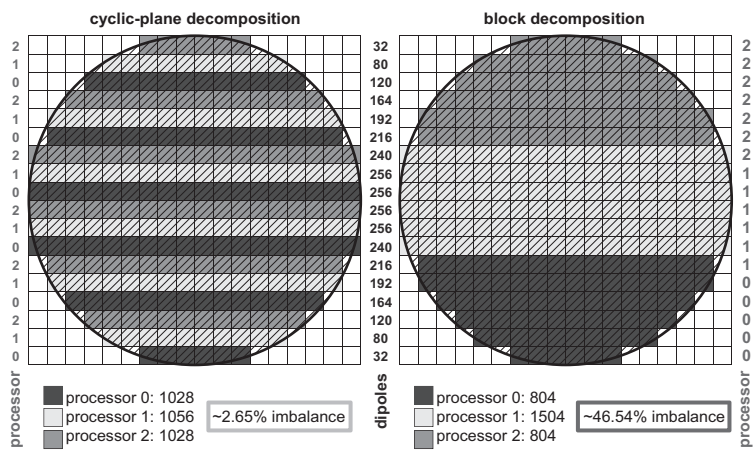}
\caption{\label{fig2}Load balance comparison for block and `cyclic-plane' decomposition for a simplistic spherical target based on an $18\!\times\!18\!\times\!18$ computational grid.}
\end{center}
\end{figure}

For the distributed-memory version, custom `cyclic-plane' domain decomposition and parallel memory allocation schemes have been developed, which, apart from permitting the efficient location of, and access to, pertinent data, also provide a decrease in memory per node that is directly proportional to the number of available nodes and inherent `to-the-nearest-plane' load balancing. This is an important point. For the majority of cases, targets do not have a constant number of occupied sites per plane. Since the custom-built iterative solution schemes only store information corresponding to occupied sites, `cyclic-plane' decomposition provides a much improved data distribution across the available nodes. For example, take the previously discussed spherical target based on an $18\!\times\!18\!\times\!18$ computational grid, in which only $3112$ of the $5832$ lattice sites are occupied. If this computational domain is `block' decomposed across three nodes, then even for this simplistic case, the imbalance of $\sim47\%$ in the data distribution dwarfs the negligible $\sim3\%$ imbalance for the `cyclic-plane' decomposition, see Fig.~\ref{fig2}. It is conceded that by `block' decomposing the constituent planes non-uniformly among the three nodes, so that node `0' is assigned the first seven planes, node `1' is assigned the four middle 4 planes, and node `2' is assigned the remaining seven planes, the load imbalance is reduced to a mere $2\%$. However, this requires that the domain decomposition and parallel memory allocation algorithms be aware of, and be able to take account of, the user-defined and essentially arbitrary target geometry. In practice, this has not been the case.

\par
To facilitate the distributed DFTs and iDFT required by the new kernel, the new framework incorporates several in-place local and distributed transpose algorithms, which have all been custom-designed to support the `cyclic-plane' decomposition. They are all capable of treating completely arbitrary systems decomposed across an arbitrary number of nodes. Please note that, unlike the serial and OpenMP versions, which only store section `1' in Fig.~\ref{fig6}, for each of the six independent tensor components, the MPI version also stores an extra octant. Rather than storing a $\left[K\times J\times P\right]$ array for each of the six components, the MPI version stores a $\left[K\times J^{\prime\prime}\times P\right]$ array, $K$ in the ${\bf {\hat x}}$ direction, $J^{\prime\prime}$ in the ${\bf {\hat y}}$ direction, and $P$ in the ${\bf {\hat z}}$ direction. This is to facilitate the distributed element-wise FD multiplication, see Eqs.~\ref{Q0eqn}$\to$\ref{Q2eqn} in Appendix \ref{origdftalgor}, that is required by the DFT kernel. This extra memory requirement is not an issue as the linear decrease in the memory per node, afforded by the parallel memory allocation, makes the extra storage inconsequential.

\section{Benchmarking}\label{benchmark}

\par
On top of the architectural specific nature of the implementations, each also exists in float, double and long double precision. For all benchmarks, double precision was used and level three and inter-procedural optimisations were used for the compilations, i.e., `-O3 -ipo' for the Intel compilers and `-O3 -ipa' for the Pathscale compilers. Furthermore, to standardise the results, all benchmarks were run using the same set of arbitrarily chosen physical parameters, {\sc SPHERE}, incident wavelength $\lambda=3.175\:\mu{\rm m}$, incident polarisation ${\rm {\hat e}_{0}}={\rm {\hat x}}$, effective radius (radius of a sphere of equal volume) $a=0.5\:\mu{\rm m}$, and complex refractive index $m=1.63631+0.372i$\footnotemark[5]\footnotetext[5]{Complex refractive index of ice for $\lambda=3.175\:\mu{\rm m}$ from the REFICE routine, originally developed by Warren in 1984 \cite{warren1984}, and subsequently improved by Warren, Gao and Wiscombe in 1995 (Unpublished, source code available in the PyARTS package, a python package developed to compliment the Atmospheric Radiative Transfer Simulator - ARTS. Available from the Satellite Atmospheric Science Group, http://www.sat.ltu.se/arts/tools).}. The following list details both the particular hardware and software configurations of each of the benchmarking systems. For each benchmark, the relevant architecture will be referenced simply by the associated tag, FRANKLIN, NEMO or WALTON.

\begin{itemize}
\item {\bf FRANKLIN: } Shared-memory, SGI Altix $3700$, Red Hat Enterprise Linux AS release $3$, IA-$64$, $24\times{\rm Intel}$ Itanium $2$, $1.5\:{\rm GHz}$, $6 \:{\rm MB}$ cache, $96\:{\rm GB}$ RAM, Intel Compilers $9.1$, FFTW $3.1.2$, Intel MKL $8.1.014$.
\item {\bf NEMO: } Shared-memory, HP xw9300 Workstation, openSUSE $10.2$, x86\_64, $2\times{\rm AMD}$ Dual-Core Opteron $275$, $2.2\:{\rm GHz}$, $2\:{\rm MB}$($2\times1\:{\rm MB}$) cache, $16\:{\rm GB}$ RAM, Intel Compilers $9.1$, FFTW $3.1.2$, MPICH2 $1.0.5p4$.
\item {\bf WALTON\footnotemark[6]\footnotetext[6]{ICHEC distributed cluster, Irish Centre for High-End Computing, http://www.ichec.ie}: } Distributed-memory, IBM cluster 1350, SUSE LINUX Enterprise Server $9$, x86\_64, $476$ compute nodes, $952\times{\rm AMD}$ Opteron Single-Core Opteron $250$, $2.4\:{\rm GHz}$, $1\:{\rm MB}$ cache, $412$ have $4\:{\rm GB}$ RAM per node, $64$ have $8\:{\rm GB}$ RAM per node, QLogic PathScale Compiler Suite, Version $3.0$, FFTW $3.1.2$, MPICH2 $1.0.2$.
\end{itemize}

\par
\begin{figure}[p]
\begin{center}
\includegraphics[scale=0.49]{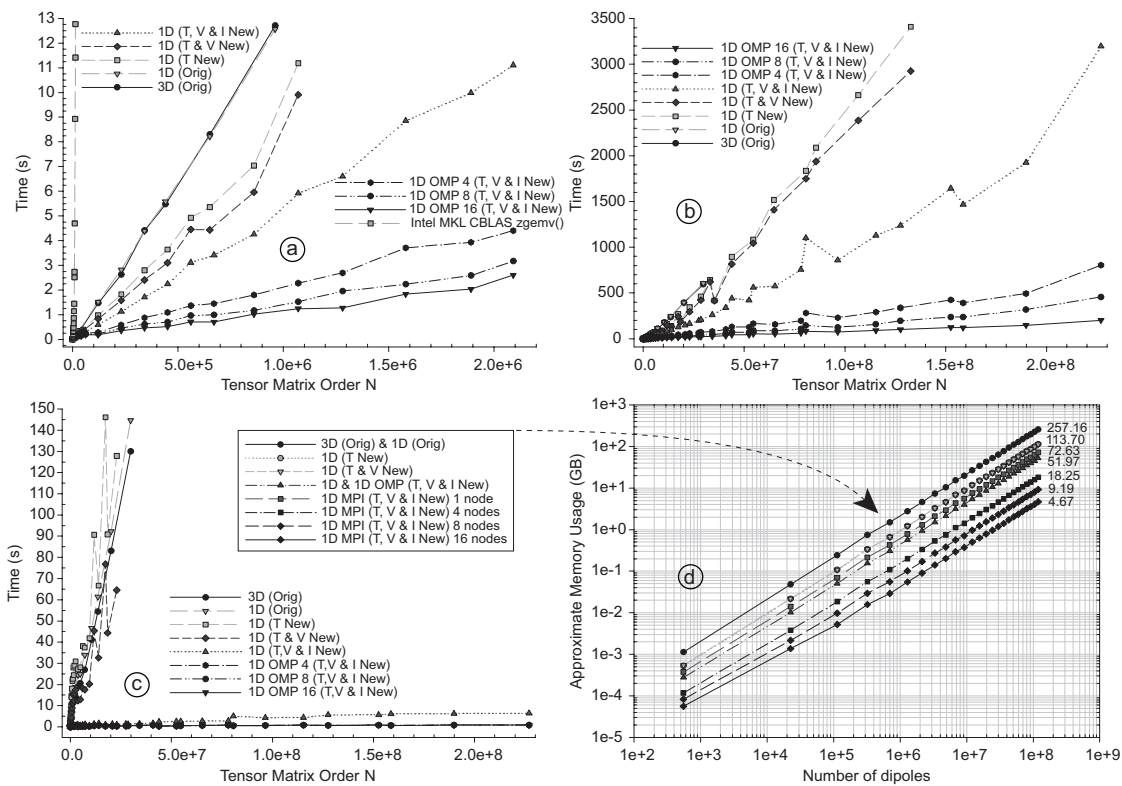}
\caption{\label{fig3} Time and memory required to compute the matrix-vector product on FRANKLIN. \small\emph{N.B. The label `Tensor Matrix Order $N$' refers to a matrix of size $N$, where the elements are tensors, i.e., the matrix is actually $3N\times3N$.} (a) Short-range timings; (b) Long-range timings; (c) FFTW DFT plan creation timings; (d) Approximate memory requirements. {\bf 3D (Orig):} Original 3D DFT algorithm for all DFTs \& iDFTs. {\bf 1D (Orig):} Original algorithm using 1D DFTS. {\bf 1D (T New):} Hybrid system, new algorithm for the tensor DFTs and the vector DFTs \& iDFTs are as in 1D (Orig). {\bf 1D (T \& V New):} Hybrid system, new algorithm for all tensor \& vector DFTS and only the iDFTs are as in 1D (Orig). {\bf 1D (T, V \& I New):} Complete new algorithm for all DFTs \& iDFTs. {\bf 1D OMP \# (T, V \& I New):} New OpenMP algorithm using `\#' threads. {\bf 1D MPI (T, V \& I New) \# nodes:} New MPI algorithm using `\#' nodes. {\bf Intel MKL CBLAS zgemv:} Explicit calculation using the Intel MKL\citep{intel2007}.}
\end{center}
\end{figure}

To affirm the benefits ascribed to the new DFT kernel, the constituent optimisations have been applied sequentially, on FRANKLIN, to show their respective contribution to the cumulative effect. Fig.~\ref{fig3}(a) shows the time to compute the matrix-vector product for the original, various hybrid, and new algorithms in a regime that allows comparison with the explicit matrix-vector multiplication. Fig.~\ref{fig3}(b) shows long-range timings, where the algorithms have been applied, within their aptitude, to much larger systems. For both the short-range and long-range cases, the timings for the {\bf 3D (Orig)} and {\bf 1D (Orig)} algorithms are essentially identical. Since the optimisation of the kernel is predicated on the concept of removing superfluous transforms by performing the requisite 3D DFTs as ensembles of 1D transforms, this congruency is crucial as it shows that there is no computational overhead associated with the migration.

\par
For the short-range case, even though the explicit matrix-vector multiplication was performed with the highly optimised Intel MKL \citep{intel2007}, it is obvious that the sheer size of the systems involved negate its use almost immediately. Furthermore, for both plots, the sequential application of the components of the new algorithm is clearly discernible. To put these improvements in perspective, in comparison to the original algorithm, the new algorithm requires approximately $60\%$ less transforms in order to carry out the matrix-vector product, which has a comparably profound influence on the timings.

\par
In the long-range case, the comparison of the original and new algorithms is limited to the nethermost region of the plot. This is due to the prohibitive memory requirements associated with the original algorithm. Note that even for extremely large systems, the new algorithm is notably faster than the original algorithm is for a system that is just a fraction of the size. Both the short-range and long-range cases also include timings for the OpenMP-based kernel across 4, 8 \& 16 threads, which unequivocally show the substantial performance gains that can be obtained from shared-memory parallelisation.

\par
An ancillary computational benefit, as shown in Fig.~\ref{fig3}(c), is that the new formulation only requires the creation of a few 1D DFT plans, rather than full 3D DFT plans. This is especially beneficial if using FFTW \citep{fftw2,fftw3} with FFTW\_MEASURE to compute the transforms, which instructs FFTW to run and measure the execution time of several DFTs in order to find the best way to compute the transform. As a result, initialisation times become negligible.

\par
In comparison to the original algorithm, the new algorithm requires over $80\%$ less memory to store the tensor and vector components. This is evident from the approximate memory requirement presented in Fig.~\ref{fig3}(d), where, for the maximum benchmarked size, corresponding to $K\!=\!J\!=\!P\!=\!610$, although the original 3D DFT algorithm would require over $257{\rm GB}$, the new algorithm and its OpenMP counterpart only require $52{\rm GB}$ to perform the same calculation. This figure also incorporates the memory requirements for the distributed MPI-based implementation, showing the reduction in memory per node that is only limited by the number of nodes available.

\par
\begin{figure}[p]
\begin{center}
\includegraphics[scale=0.5]{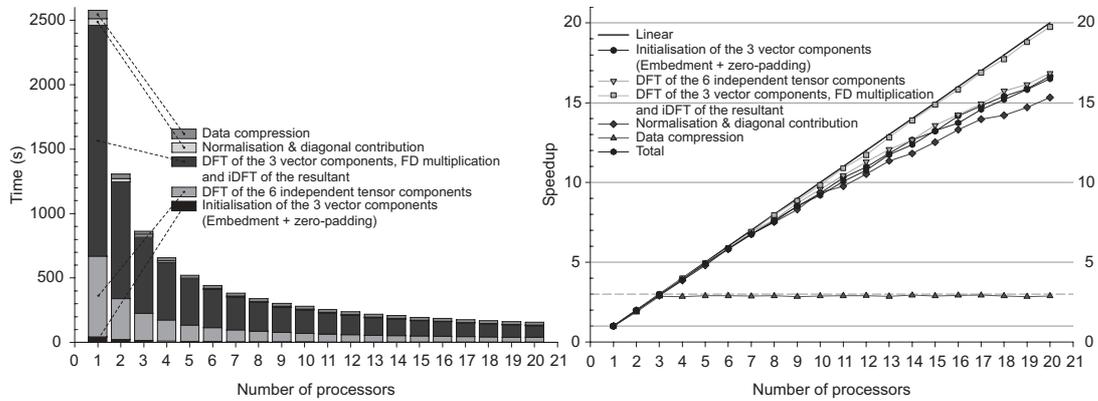}
\caption{\label{fig4}OpenMP scalability and timing run on FRANKLIN for a spherical target with $K\!=\!J\!=\!P\!=\!600$ and $N_{d}\!=\!113,089,912$ occupied sites.}
\end{center}
\end{figure}

To ascertain the performance and scalability of the OpenMP-based shared-memory implementation, a large-scale benchmark was performed across $1,2,\ldots,20$ threads on FRANKLIN, see Fig.~\ref{fig4}. The system size was chosen to be $K\!=\!J\!=\!P\!=600$, which produced a computational domain with $N\!=\!216\times\!10^{6}$ sites in total and a representation of the spherical target with $N_{d}\!=\!113,098,912$ occupied sites. In terms of memory usage, this system required $49{\rm GB}$ for 1 thread, with a linear increase of approximately $66{\rm MB}$ per thread to approximately $50{\rm GB}$ for 20 threads. The extra memory is for the thread private scratch spaces that are required to ensure thread safety. Although this may sound like large amount of memory, in comparison, for the same system, the original algorithm, which is limited to the serial environment, would require approximately $242{\rm GB}$. The time for the matrix-vector product is reduced from $2576\:{\rm s}$ for 1 thread to just $156\:{\rm s}$ using 20 threads, a speedup of approximately $16.5$. All components show excellent scalability, except of course, the aforementioned data compression algorithm, see \S\ref{parallelimp}, which is not an issue as it constitutes only a negligible fraction of the DFT kernel's runtime.

\par
To test the performance and scalability of the MPI-based distributed-memory implementation, the same $K\!=\!J\!=\!P\!=\!600$ system as for the OpenMP shared-memory benchmark, was run on WALTON, see Fig.~\ref{fig4}. Here, due to the fact that each node has only $4{\rm GB}$ of physical memory, the benchmark was initialised across 32 nodes and then scaled up to 64 nodes, requiring only $2.2{\rm GB}$ per node across 32 nodes, and just $1.2{\rm GB}$ per node across 64 nodes. In terms of performance, the timings and scalability are extremely good. The calculation took $108\:{\rm s}$ across 32 nodes and $55\:{\rm s}$ across 64 nodes, a speedup of almost $2$. To put this in perspective, for the OpenMP-based benchmark, the same system required $2576\:{\rm s}$ using 1 thread and $156\:{\rm s}$ using 20 threads. In terms of these results, there are several noteworthy points. Firstly, the total time is quoted as the sum of the computation and communication times. The total communication time includes the time for the forward transpositions of both the six independent tensor components and three zero-padded vector components, and also includes the time for the reverse transposition of the resultant of the FD element-wise multiplication of the components. These transpositions are essentially block events that are controlled by quite complex custom-built transposition algorithms. Even with the prescribed optimisations, due to the enormity of the tensor and vector component arrays, `in-place' transposition is an absolute necessity. This complication is compounded by the custom-built parallel-memory allocation and `cyclic-plane' domain decomposition schemes that provide a decrease in memory per node that is directly proportional to the number of active nodes and also provide inherent load balancing. In order to facilitate efficient `in-place' transposition of the component arrays, the transposition algorithms employ the pre-computation of local `in-place' transposes in order to simplify and streamline the host algorithm and subsequent data communication. Unfortunately, at present this negates the possibility of overlapping some of the communication with the computation. Secondly, although all communication timings are marginally sub-linear, all computation timings exhibit super-linear speedups. Upon consulting with the ICHEC\footnotemark[7]\footnotetext[7]{Irish Centre for High-End Computing, http://www.ichec.ie} technical systems team, the peculiarity is believed to be an artifact of the initial benchmark. Due to the novel parallel memory allocation algorithms in the distributed-memory implementation of the OpenDDA framework, the memory footprint is decomposed across the active nodes. However, it is believed that the chosen system size and associated memory requirements were fractionally too large, causing slightly degraded performance due to swapping. Since all subsequent speedups are calculated with respect to this initial benchmark, and since the parallel memory allocation and domain decomposition schemes would have solved this allocation/swapping issue as the number of active nodes increased above 32, the subsequent timings would appear super-linear. This is not considered to be a cause for concern as this only introduces a global shift and it is to overall trend that is of importance here.

\par
To test the performance of the full OpenDDA framework, a benchmark was run on NEMO with $K\!=\!J\!=\!P\!=200$, see Table.~\ref{table3}. This produced a computational domain with $N\!=\!8\times\!10^{6}$ sites in total and a representation of the spherical target with $N_{d}\!=\!4,188,896$ occupied sites. The convergence tolerance was taken as the default value of $1\!\times\!10^{-10}$, the maximum number of iterations was set to $1\!\times\!10^{4}$, the initial guess was set to zero, and Point-Jacobi preconditioning was used. The performance and scalability of the full OpenDDA framework is extremely good. For this system, it is the simpler schemes that tend to provide the best performance, however, since convergence tends to be slower for larger refractive index values \citep{draine1988}, as the number of iterations increases, the enhanced stability of the more involved iterative methods may prove beneficial, and ultimately, may provide superior convergence. Although it is duly noted that due to the necessity of distributed transposes, the performance of the MPI version will always lag behind that of the OpenMP version, the slight deficit in performance is offset by the decrease in memory that is directly proportional to the number of nodes available.

\par
To put the new custom-built OpenDDA framework in context, several comparative benchmarks were performed between the various architecture specific implementations of OpenDDA, and ADDA; the fastest and most computationally efficient DDA implementation \citep{penttila2007}. Since the number of iterations to reach convergence is dependent on the user-defined convergence tolerance and on the way in which the stopping criteria are defined within the specific implementation, the comparison focusses, not on the number of iterations, but on the time per iteration. For ADDA, the reason that only the total time for the initialisation of the tensor components is given is that ADDA performs the initialisation of the tensor components as a single conjoint operation, i.e., the computation of the components and the DFTs of these components are perform together. In terms of the quoted memory requirements, for the OpenDDA variants, the approximate memory requirements are obtained from the in-built estimation routines. To obtain an estimate for the approximate memory requirements for ADDA, only `large' arrays pertaining to the iterative solution have been included. Furthermore, some of the `large' arrays have been excluded from consideration, as in ADDA, not all relevant `large' arrays coexist, and thus, some do not contribute to the cumulative memory requirements. For the shared-memory comparison, for ${\rm BICG}_{{\rm sym}}$, the memory estimations for ADDA include, $[$\textit{xvec}, \textit{rvec}, \textit{pvec}, \textit{Einc}, \textit{Avecbuffer}, \textit{Dmatrix}, \textit{Xmatrix}, \textit{slices} \& \textit{slices\_tr}$]$, and for ${\rm BiCGSTAB}$ \& ${\rm QMR}_{{\rm sym}}$, the estimation also includes $[$\textit{vec1}, \textit{vec2} \& \textit{vec3}$]$. For the distributed-memory comparison, the estimations also include the contributions from ADDA's MPI communication buffers, $[$\textit{BT\_buffer} \& \textit{BT\_rbuffer}$]$, and to look at the efficiency of the decomposition of the associated data across the active nodes, apart from the total memory requirements, the results also give the maximum memory requirement per node.

\par
Table \ref{tablenemoadda200} shows the results for the shared-memory comparative benchmark run on NEMO using ${\rm BiCGSTAB}$, ${\rm BICG}_{{\rm sym}}$ \& ${\rm QMR}_{{\rm sym}}$, for a spherical target with $K\!=\!J\!=\!P\!=\!200$, $N\!=\!8\times\!10^{6}$, and $N_{d}\!=\!4,188,896$. Please note that although ADDA does employ Intel vectorisation \citep{intel2004} within its basic linear algebra routines, it does incorporate generic OpenMP shared-memory parallelisation. Similarly, Table \ref{tablewaltonadda} shows the results for the distributed-memory comparative benchmark run on WALTON using ${\rm BiCGSTAB}$, ${\rm BICG}_{{\rm sym}}$ \& ${\rm QMR}_{{\rm sym}}$, across $4$, $8$, $16$, $32$ \& $64$ nodes, where the relevant system sizes for each of the constituent benchmarks are quoted in Table \ref{tablesyssizewaltadda}. In all instances, the new OpenDDA performs extremely favourably. It is worth noting that the custom `cyclic-plane' decomposition within the OpenDDA framework allows for an extremely even distribution of the data across the active nodes. Even though, for the comparison across 32 nodes in Table \ref{tablewaltonadda}, OpenDDA requires marginally more memory in total, the maximum memory required by any individual node is actually less.

\par
Finally, as a quick verification of the correct operation of the new OpenDDA framework, a comparison of the extinction, scattering \& absorption efficiencies, $Q_{ext}$, $Q_{sca}$ \& $Q_{abs}$, for the well proven output of the widely used DDSCAT code and the three architectural variants of OpenDDA was performed on NEMO, see Table \ref{tableopenddavalid}. All are in good agreement.

\section{Conclusions}\label{conclusions}

\par
Within the new framework, the bit-fielding of the storage associated with the spatial and compositional specification of the target, not only provides significant reduction in the associated integer storage, but also facilitates the creation of a versatile and efficient distributed-memory implementation.

\par
Although comparable codes, such as ADDA, have pioneered and employ similar optimisations, OpenDDA incorporates a custom-built highly-optimised DFT kernel that performs the requisite 3D DFTs of both the tensor and vector, which are instrumental in the efficient calculation of the matrix-vector products, as ensembles of 1D transforms. By doing so, a significant fraction of the constituent 1D transforms, which are superfluous, can be neglected. In comparison to the ``out-of-the-box'' algorithm based on 3D DFTs, OpenDDA requires approximately $60\%$ less transforms and over $80\%$ less memory in order to carry out the matrix-vector products. Furthermore, the optimisations permit parallel techniques to be employed to further enhance the performance. Both the resulting OpenMP-based shared-memory and MPI-based distributed-memory implementations possess extremely good scalability and associated performance. On top of these improvements, due to the fact that, computationally, the new framework only requires the creation of a limited number of 1D DFT plans, rather than the full 3D DFT plans required by the original algorithm, the initialisation times, which can be significant, become completely negligible. The new framework also incorporates a relatively large selection of iterative methods that have been custom-built in serial, OpenMP and MPI to integrate seamlessly with the various architecturally specific DFT kernels. 

\par
Within the distributed-memory implementation of the new framework, the novel parallel-memory allocation and domain decomposition schemes provide a decrease in memory per node that is directly proportional to the number of nodes and also provide inherent load balancing. Due to the fact that the new framework only stores information corresponding to occupied lattice sites, depending on the target being simulated, there can be quite an imbalance in the number of occupied sites per plane. However, the new framework employs custom-designed `cyclic-plane' decomposition, and as a result, boasts an extremely well-balanced data distribution across the active nodes.

\par
The framework performs extremely favourably across all architectural variants, with the novel and unique OpenMP-based implementation providing unprecedented parallel performance on shared-memory architectures.

\section*{Acknowledgements}
{This work was supported by the CosmoGrid Project\footnotemark[8]\footnotetext[8]{CosmoGrid Project, http://www.cosmogrid.ie}, funded under the Programme for Research in Third Level Institutions (PRTLI) administered by the Irish Higher Education Authority under the National Development Plan and with partial support from the European Regional Development Fund. In conclusion, the authors would also like to thank the Irish Centre for High-End Computing (ICHEC)\footnotemark[9]\footnotetext[9]{Irish Centre for High-End Computing, http://www.ichec.ie} and their dedicated Systems Team.}

\bibliographystyle{chicago}
\bibliography{databib}

\appendix

\section{Original DFT algorithm}\label{origdftalgor}

\par
This appendix concisely summarises the conventional application of DFTs to the solution of complex-symmetric level-three tensor-Toeplitz matrix-vector products in the context of the DDA, as prescribed by \cite{goodman1991}.

\par
The extension of the level-one formulation described at the beginning of \S\ref{kernelopt} to the treatment of level-three systems, ${}^{{\scscs[N]}}{\rm y}={}^{{\scscs[P][J][K]}}{\rm T}{}^{{\scscs[N]}}{\rm x}$, is obtained by performing the T-to-c conversion on each of the three levels of ${}^{{\scscs[P][J][K]}}{\rm T}$ just as it was performed for the level-one system, resulting in a level-three circulant, ${}^{{\scscs[P^{\prime}][J^{\prime}][K^{\prime}]}}{\rm C}$, where $K^{\prime}=2K-1,J^{\prime}=2J-1,P^{\prime}=2P-1$.

\vspace{-1em}
\begin{eqnarray}
{}^{{\scscs[P^{\prime}][J^{\prime}][K^{\prime}]}}{\rm C}&\!\!=\!\!&\sum\limits_{k=0}^{P^{\prime}-1}\left[{}^{{\scscs[P^{\prime}]}}{\psi}^{k}\right]\varotimes\left[^{{\scscs[J^{\prime}][K^{\prime}]}}{\rm C}_{k}\right] \nn \\
{}^{{\scscs[J^{\prime}][K^{\prime}]}}{\rm C}_{k}&\!\!=\!\!&\sum\limits_{j=0}^{J^{\prime}-1}\left[{}^{{\scscs[J^{\prime}]}}{\psi}^{j}\right]\varotimes\left[^{{\scscs[K^{\prime}]}}{\rm C}_{k,j}\right] \nn \\
{}^{{\scscs[K^{\prime}]}}{\rm C}_{k,j}&\!\!=\!\!&\sum\limits_{i=0}^{K^{\prime}-1}{\rm C}_{k,j,i}\left[{}^{{\scscs[K^{\prime}]}}{\psi}^{i}\right] \nn
\end{eqnarray}

\ni
Here, each of the $^{{\scscs[J^{\prime}][K^{\prime}]}}{\rm C}_{k}$ is a level-two circulant, each of the $^{{\scscs[K^{\prime}]}}{\rm C}_{k,j}$ is a level-one circulant, the symbol `$\varotimes$' denotes the Kronecker tensor product, and ${}^{{\scscs[n]}}{\psi}$ is the downward-shift permutation matrix of order $[n]$ \citep{vanloan1992}. The arbitrary vector must also be zero-padded to negate the contribution for the circulant extensions on all three levels, ${}^{{\scscs[N]}}{\rm x}\to{}^{{\scscs[N^{\prime}]}}{\rm x}$, where $[N^{\prime}]=[K^{\prime}\times J^{\prime}\times P^{\prime}]$. The complimentary Fourier structure which diagonalises this level-three system is acquired from the hierarchical diagonalisation by the Fourier matrices, ${}^{{\scscs[P^{\prime}]}}{\rm F}$, ${}^{{\scscs[J^{\prime}]}}{\rm F}$ and ${}^{{\scscs[K^{\prime}]}}{\rm F}$, i.e., ${}^{{\scscs[P^{\prime}][J^{\prime}][K^{\prime}]}}{\rm F}={}^{{\scscs[P^{\prime}]}}{\rm F}\varotimes{}^{{\scscs[J^{\prime}]}}{\rm F}\varotimes{}^{{\scscs[K^{\prime}]}}{\rm F}$ \citep{davis1994}. As a result, the level-three matrix-vector product, ${}^{{\scscs[N^{\prime}]}}{\rm y}={}^{{\scscs[P^{\prime}][J^{\prime}][K^{\prime}]}}{\rm C}{}^{{\scscs[N^{\prime}]}}{\rm x}$, can be carried out as an element-wise FD multiplication, where ${}^{{\scscs[N^{\prime}]}}{\rm c}$ is the first column of ${}^{{\scscs[P^{\prime}][J^{\prime}][K^{\prime}]}}{\rm C}$.

\vspace{-1em}
\begin{eqnarray}
{}^{{\scscs[N^{\prime}]}}{\rm y}={}^{{\scscs[P^{\prime}][J^{\prime}][K^{\prime}]}}{\rm F}^{-1}\left[\sqrt{P^{\prime}J^{\prime}K^{\prime}}\left[{}^{{\scscs[P^{\prime}][J^{\prime}][K^{\prime}]}}{\rm F}\right]{}^{{\scscs[N^{\prime}]}}{\rm c}\right].\left[{}^{{\scscs[P^{\prime}][J^{\prime}][K^{\prime}]}}{\rm F}{}^{{\scscs[N^{\prime}]}}{\rm x}\right] \label{threelevelmatvecprodelementwise}
\end{eqnarray}

\par
Furthermore, using the realisation that ${}^{{\scscs[P^{\prime}][J^{\prime}][K^{\prime}]}}{\rm F}\equiv$ 3D DFT allows Eq.~\ref{threelevelmatvecprodelementwise} to be reformulated in terms of the 3D DFT. To facilitate this, before the aforementioned T-to-c conversions and zero-padding takes place, the first column of each of the six independent tensor component matrices, ${}^{{\scscs[N]}}{\rm t0},\ldots,{}^{{\scscs[N]}}{\rm t5}$, and the three vector components, ${}^{{\scscs[N]}}{\rm x0},{}^{{\scscs[N]}}{\rm x1},{}^{{\scscs[N]}}{\rm x2}$, must be arranged into separate 3D structures, ${}^{{\scscs[N]}}{\rm t0{\scs3D}},\ldots,{}^{{\scscs[N]}}{\rm t5{\scs3D}}$ and ${}^{{\scscs[N]}}{\rm x0{\scs 3D}},{}^{{\scscs[N]}}{\rm x1{\scs 3D}},{}^{{\scscs[N]}}{\rm x2{\scs 3D}}$, respectively. For example, the transformations ${}^{{\scscs[N]}}{\rm t0}\rightarrow{}^{{\scscs[N]}}{\rm t0{\scs3D}}$ and ${}^{{\scscs[N]}}{\rm x0}\rightarrow{}^{{\scscs[N]}}{\rm x0{\scs 3D}}$ are governed by ${}^{{\scscs[N]}}{\rm t0{\scs3D}}_{i,j,k}={}^{{\scscs[N]}}{\rm t0}_{kJK+jK+i}$ and ${}^{{\scscs[N]}}{\rm x0{\scs 3D}}_{i,j,k}={}^{{\scscs[N]}}{\rm x0}_{kJK+jK+i}$, $i=0,\ldots,K-1,j=0,\ldots,J-1,k=0,\ldots,P-1$, respectively.

\par
Once in 3D form, to facilitate the DFT applicability, each of the six tensor-component arrays must undergo T-to-c conversion, e.g., ${}^{{\scscs[N]}}{\rm t0{\scs3D}}\to{}^{{\scscs[N^{\prime}]}}{\rm c0{\scs3D}}$, and the three vector-component arrays must be zero-padded, e.g., ${}^{{\scscs[N]}}{\rm x0{\scs 3D}}\rightarrow{}^{{\scscs[N^{\prime}]}}{\rm x0{\scs 3D}}$. For a level-three system, the T-to-c conversion involves taking each tensor-component array in turn, denoted by section `$1$' in Fig.~\ref{fig6}, treating each line in the ${\bf {\hat x}}$, ${\bf {\hat y}}$ \& ${\bf {\hat z}}$ directions as the first column of a level-one Toeplitz matrix, and mirroring it, excluding the initial entry, across its last entry. The mirrored elements must also by multiplied by the appropriate entry from Table~\ref{table1}. The reason for this is that for a general T-to-c conversion, one would usually append the first row, excluding the initial entry, in reverse order. However, since the nine tensor components are from the dyadic product in Eq.~\ref{dyadicprod}, the Toeplitz matrix is either symmetric or non-symmetric by sign only. Hence, it is possible to just mirror the column itself and, if necessary, multiply the mirrored entries by `$-1$'. The ${\bf {\hat x}}$ conversion creates section `$2$', the ${\bf {\hat y}}$ conversion creates section `$3$', and the ${\bf {\hat z}}$ conversion creates section `$4$' in Fig.~\ref{fig6}. The zero-padding of the three vector-component arrays is much simpler. Each component is trivially embedded in the $[0\rightarrow K-1]\times[0\rightarrow J-1]\times[0\rightarrow P-1]$ octant of a $K^{\prime}\times J^{\prime}\times P^{\prime}$ array of zeroes, i.e., it would occupy section `$1$' in Fig.~\ref{fig6}, and the rest of the array would be identically zero.

\par
The algorithm requires the DFT of the six tensor-component arrays, ${}^{{\scscs[N^{\prime}]}}{\rm c0{\scs 3D}},\ldots,{}^{{\scscs[N^{\prime}]}}{\rm c5{\scs 3D}}$, and the three vector-component arrays, ${}^{{\scscs[N^{\prime}]}}{\rm x0{\scs 3D}},{}^{{\scscs[N^{\prime}]}}{\rm x1{\scs 3D}},{}^{{\scscs[N^{\prime}]}}{\rm x2{\scs 3D}}$, yielding ${\rm Y1},{\rm Y2},\ldots,{\rm Y5}$ and ${\rm X0},{\rm X1},{\rm X2}$, respectively. Note that for the practical application of the theory, as with DDSCAT \citep{draine2004} and ADDA \citep{yurkin2008}, for maximum performance, if necessary, the transformations described above are all zero-padded to ensure that the final dimensions of the component arrays permit the use of fast DFT algorithms, i.e., for FFTW \citep{fftw2,fftw3}, a dimension which satisfies $2^{a}3^{b}5^{c}7^{d}11^{e}13^{f}$, where $a,b,c,d$ are arbitrary and $(e+f\le1)$. These are illustrated in Fig.~\ref{fig6} as $kf$, $jf$, and $pf$, and increase the dimensions to $K^{\prime\prime}=K^{\prime}+kf=2K-1+kf$, $J^{\prime\prime}=J^{\prime}+jf=2J-1+jf$, and $P^{\prime\prime}=P^{\prime}+pf=2P-1+pf$. Subsequently, the contribution from each of the individual components, to the full circulant matrix-vector product, is obtained via the element-wise tensor-vector product of their transforms.

\vspace{-1em}
\begin{eqnarray}
{\rm Q0}_{i,j,k}&=&{\rm Y0}_{i,j,k}{\rm X0}_{i,j,k}+{\rm Y1}_{i,j,k}{\rm X1}_{i,j,k}+{\rm Y2}_{i,j,k}{\rm X2}_{i,j,k} \label{Q0eqn} \\
{\rm Q1}_{i,j,k}&=&{\rm Y1}_{i,j,k}{\rm X0}_{i,j,k}+{\rm Y3}_{i,j,k}{\rm X1}_{i,j,k}+{\rm Y4}_{i,j,k}{\rm X2}_{i,j,k} \label{Q1eqn} \\
{\rm Q2}_{i,j,k}&=&{\rm Y2}_{i,j,k}{\rm X0}_{i,j,k}+{\rm Y4}_{i,j,k}{\rm X1}_{i,j,k}+{\rm Y5}_{i,j,k}{\rm X2}_{i,j,k} \label{Q2eqn} \\
i&=&0,\ldots,K^{\prime}-1,j=0,\ldots,J^{\prime}-1,k=0,\ldots,P^{\prime}-1 \nn
\end{eqnarray}

\par
The inverse DFT (iDFT) of ${\rm Q0},{\rm Q1}$ \& ${\rm Q2}$ provides ${\rm y0{\scs 3D}},{\rm y1{\scs 3D}}$ \& ${\rm y2{\scs 3D}}$, which are the resultant of Eq.~\ref{threelevelmatvecprodelementwise}, in 3D form, and the $x,y,z$ components of the original tensor-Toeplitz matrix-vector product, ${}^{{\scscs[N]}}{\rm y}={}^{{\scscs[P][J][K]}}{\rm T}{}^{{\scscs[N]}}{\rm x}$, are obtained by extracting the $[0\rightarrow K-1]\times[0\rightarrow J-1]\times[0\rightarrow P-1]$ octant of ${\rm y0{\scs 3D}}$, ${\rm y1{\scs 3D}}$ \& ${\rm y2{\scs 3D}}$, respectively, i.e., `$1$' in Fig.~\ref{fig6}. This reduces the complexity from $O([3N_{d}]^{2})$ operations and storage to $O(12N^{\prime}\ln N^{\prime})$ and $O(9N^{\prime})$, respectively.

\begin{table*}[p]
\begin{center}
\caption{\label{table1} Signs to account for the non-symmetries in sign only for the T-to-c conversions of the six independent tensor components from the dyadic product in Eq.~\ref{dyadicprod}.}
\begin{tabular}[p]{|c|cccccc|}
\cline{2-7}
\multicolumn{1}{c|}{}&$xx$&$xy$&$xz$&$yy$&$yz$&$zz$\\
\hline
$x$ DFT&+1&-1&-1&+1&+1&+1\\
$y$ DFT&+1&-1&+1&+1&-1&+1\\
$z$ DFT&+1&+1&-1&+1&-1&+1\\
\hline
\end{tabular}
\end{center}
\end{table*}

\begin{table*}[p]
\begin{center}
\caption{\label{table2} Complexity estimations for the original and new algorithms.}
\begin{tabular}[p]{|c|c|c|}
\cline{2-3}
\multicolumn{1}{c|}{$\scs\rm\bf{{COMPUTATION}}$}&${\scs\rm\bf{{Original\:Algorithm}}}$&\multicolumn{1}{c|}{$\scs\rm\bf{{New\:Algorithm}}$} \\
\hline
${\scs {\rm Y0}\rightarrow{\rm Y5}:{\bf {\hat x}}}$&$O(6N^{\prime\prime}\ln K^{\prime\prime})$&$O(6JPK^{\prime\prime}\ln K^{\prime\prime})$ \\
${\scs {\rm Y0}\rightarrow{\rm Y5}:{\bf {\hat y}}}$&$O(6N^{\prime\prime}\ln J^{\prime\prime})$&$O(6\left[K+kf\right]PJ^{\prime\prime}\ln J^{\prime\prime})$ \\
${\scs {\rm Y0}\rightarrow{\rm Y5}:{\bf {\hat z}}}$&$O(6N^{\prime\prime}\ln P^{\prime\prime})$&$O(6\left[K+kf\right]\left[J+jf\right]P^{\prime\prime}\ln P^{\prime\prime})$ \\
\hline
${\scs {\rm X0}\rightarrow{\rm X2}:{\bf {\hat x}}}$&$O(3N^{\prime\prime}\ln K^{\prime\prime})$&$O(3J^{\prime\prime}PK^{\prime\prime}\ln K^{\prime\prime})$ \\
${\scs {\rm X0}\rightarrow{\rm X2}:{\bf {\hat y}}}$&$O(3N^{\prime\prime}\ln J^{\prime\prime})$&$O(3KPJ^{\prime\prime}\ln J^{\prime\prime})$ \\
${\scs {\rm X0}\rightarrow{\rm X2}:{\bf {\hat z}}}$&$O(3N^{\prime\prime}\ln P^{\prime\prime})$&$O(3N^{\prime\prime}\ln P^{\prime\prime})$ \\
\hline
${\scs {\rm Q0}\rightarrow{\rm Q2}:{\bf {\hat x}}}$&$O(3N^{\prime\prime}\ln K^{\prime\prime})$&$O(3J^{\prime\prime}PK^{\prime\prime}\ln K^{\prime\prime})$ \\
${\scs {\rm Q0}\rightarrow{\rm Q2}:{\bf {\hat y}}}$&$O(3N^{\prime\prime}\ln J^{\prime\prime})$&$O(3KPJ^{\prime\prime}\ln J^{\prime\prime})$ \\
${\scs {\rm Q0}\rightarrow{\rm Q2}:{\bf {\hat z}}}$&$O(3N^{\prime\prime}\ln P^{\prime\prime})$&$O(3N^{\prime\prime}\ln P^{\prime\prime})$ \\
\hline
\multicolumn{1}{c}{}&\multicolumn{1}{c}{}&\multicolumn{1}{c}{}\\
\cline{2-3}
\multicolumn{1}{c|}{$\scs\rm\bf{{STORAGE}}$}&${\scs\rm\bf{{Original\:Algorithm}}}$&\multicolumn{1}{c|}{$\scs\rm\bf{{New\:Algorithm}}$} \\
\hline
${\scs {\rm Y0}\rightarrow{\rm Y5}}$&$O(6N^{\prime\prime})$&$O(6\left[(K+kf)\times(J+jf)\times(P+pf)\right])$ \\
${\scs {\rm X0}\rightarrow{\rm X2}}$&$O(3N^{\prime\prime})$&$O(3\left[KJ^{\prime\prime}P\right])$ \\
\hline
\end{tabular}
\end{center}
\end{table*}

\scriptsize
\begin{table*}[p]
\begin{center}
\caption{\label{table3} Performance of the full serial, OpenMP \& MPI implementations of the new framework on NEMO using 4 processors for the parallel versions. {\bf SV:} Number of starting vectors, {\bf NI:} Number of iterations, {\bf CT:} Convergence time, {\bf SU:} Speedup, {\bf Mem:} Memory required in total, {\bf Node:} Memory required per node.}
\begin{tabular}[p]{|l|c|c|c|c|c|c|c|c|c|c|c|}
\cline{4-12}
\multicolumn{3}{c|}{$\scs\rm\bf{{}}$}&\multicolumn{2}{|c|}{$\scs\rm\bf{{SERIAL}}$}&\multicolumn{3}{|c|}{$\scs\rm\bf{{OpenMP}}$}&\multicolumn{4}{|c|}{$\scs\rm\bf{{MPI}}$} \\
\hline
${\scs\rm\bf{{Scheme}}}$&${\scs\rm\bf{{SV}}}$&${\scs\rm\bf{{NI}}}$&${\scs\rm\bf{{CT}}}$&${\scs\rm\bf{{Mem}}}$&${\scs\rm\bf{{CT}}}$&${\scs\rm\bf{{SU}}}$&${\scs\rm\bf{{Mem}}}$&${\scs\rm\bf{{CT}}}$&${\scs\rm\bf{{SU}}}$&${\scs\rm\bf{{Mem}}}$&${\scs\rm\bf{{Node}}}$ \\
${\scs\rm\bf{{}}}$&${\scs\rm\bf{{}}}$&${\scs\rm\bf{{}}}$&${\scs\rm\bf{{(s)}}}$&${\scs\rm\bf{{({\rm GB})}}}$&${\scs\rm\bf{{(s)}}}$&${\scs\rm\bf{{}}}$&${\scs\rm\bf{{({\rm GB})}}}$&${\scs\rm\bf{{(s)}}}$&${\scs\rm\bf{{}}}$&${\scs\rm\bf{{({\rm GB})}}}$&${\scs\rm\bf{{({\rm GB})}}}$ \\
\hline
${\scs\rm{{bicg}}}$&${\scs{{-}}}$&${\scs{{21}}}$&${\scs{{1439.7}}}$&${\scs{{3.1}}}$&${\scs{{404.3}}}$&${\scs{{3.56}}}$&${\scs{{3.2}}}$&${\scs{{604.4}}}$&${\scs{{2.4}}}$&${\scs{{3.9}}}$&${\scs{{1.0}}}$ \\
${\scs\rm{{bicg_{sym}}}}$&${\scs{{-}}}$&${\scs{{21}}}$&${\scs{{741.5}}}$&${\scs{{2.6}}}$&${\scs{{212.7}}}$&${\scs{{3.5}}}$&${\scs{{2.6}}}$&${\scs{{306.2}}}$&${\scs{{2.4}}}$&${\scs{{3.3}}}$&${\scs{{0.8}}}$ \\
${\scs\rm{{bicgstab}}}$&${\scs{{-}}}$&${\scs{{12}}}$&${\scs{{846.6}}}$&${\scs{{3.1}}}$&${\scs{{234.7}}}$&${\scs{{3.6}}}$&${\scs{{3.2}}}$&${\scs{{353.8}}}$&${\scs{{2.4}}}$&${\scs{{3.9}}}$&${\scs{{1.0}}}$ \\
${\scs\rm{{cg}}}$&${\scs{{-}}}$&${\scs{{21}}}$&${\scs{{746.6}}}$&${\scs{{2.6}}}$&${\scs{{209.7}}}$&${\scs{{3.6}}}$&${\scs{{2.6}}}$&${\scs{{320.9}}}$&${\scs{{2.3}}}$&${\scs{{3.3}}}$&${\scs{{0.8}}}$ \\
${\scs\rm{{cgs}}}$&${\scs{{-}}}$&${\scs{{13}}}$&${\scs{{911.3}}}$&${\scs{{3.1}}}$&${\scs{{257.4}}}$&${\scs{{3.5}}}$&${\scs{{3.2}}}$&${\scs{{382.9}}}$&${\scs{{2.4}}}$&${\scs{{3.9}}}$&${\scs{{1.0}}}$ \\
${\scs\rm{{mlbicgstab}}}$&${\scs{{1}}}$&${\scs{{12}}}$&${\scs{{836.0}}}$&${\scs{{3.3}}}$&${\scs{{238.9}}}$&${\scs{{3.5}}}$&${\scs{{3.3}}}$&${\scs{{341.5}}}$&${\scs{{2.5}}}$&${\scs{{4.0}}}$&${\scs{{1.0}}}$ \\
${\scs\rm{{mlbicgstab}}}$&${\scs{{2}}}$&${\scs{{8}}}$&${\scs{{811.7}}}$&${\scs{{3.9}}}$&${\scs{{246.2}}}$&${\scs{{3.3}}}$&${\scs{{3.9}}}$&${\scs{{347.1}}}$&${\scs{{2.3}}}$&${\scs{{4.6}}}$&${\scs{{1.2}}}$ \\
${\scs\rm{{mlbicgstab}}}$&${\scs{{3}}}$&${\scs{{6}}}$&${\scs{{842.9}}}$&${\scs{{4.5}}}$&${\scs{{266.2}}}$&${\scs{{3.2}}}$&${\scs{{4.5}}}$&${\scs{{369.6}}}$&${\scs{{2.3}}}$&${\scs{{5.2}}}$&${\scs{{1.3}}}$ \\
${\scs\rm{{qmr}}}$&${\scs{{-}}}$&${\scs{{22}}}$&${\scs{{3704.6}}}$&${\scs{{3.5}}}$&${\scs{{1055.1}}}$&${\scs{{3.5}}}$&${\scs{{3.5}}}$&${\scs{{1636.4}}}$&${\scs{{2.3}}}$&${\scs{{4.3}}}$&${\scs{{1.1}}}$ \\
${\scs\rm{{qmr_{sym}}}}$&${\scs{{-}}}$&${\scs{{21}}}$&${\scs{{759.6}}}$&${\scs{{3.1}}}$&${\scs{{216.7}}}$&${\scs{{3.5}}}$&${\scs{{3.2}}}$&${\scs{{320.7}}}$&${\scs{{2.4}}}$&${\scs{{3.9}}}$&${\scs{{1.0}}}$ \\
${\scs\rm{{rbicgstab}}}$&${\scs{{1}}}$&${\scs{{14}}}$&${\scs{{996.3}}}$&${\scs{{3.1}}}$&${\scs{{284.5}}}$&${\scs{{3.5}}}$&${\scs{{3.2}}}$&${\scs{{456.0}}}$&${\scs{{2.2}}}$&${\scs{{3.9}}}$&${\scs{{1.0}}}$ \\
${\scs\rm{{rbicgstab}}}$&${\scs{{2}}}$&${\scs{{7}}}$&${\scs{{990.5}}}$&${\scs{{3.5}}}$&${\scs{{285.2}}}$&${\scs{{3.5}}}$&${\scs{{3.5}}}$&${\scs{{415.5}}}$&${\scs{{2.4}}}$&${\scs{{4.3}}}$&${\scs{{1.1}}}$ \\
${\scs\rm{{rbicgstab}}}$&${\scs{{3}}}$&${\scs{{4}}}$&${\scs{{851.1}}}$&${\scs{{3.9}}}$&${\scs{{242.2}}}$&${\scs{{3.5}}}$&${\scs{{3.9}}}$&${\scs{{357.7}}}$&${\scs{{2.4}}}$&${\scs{{4.6}}}$&${\scs{{1.2}}}$ \\
${\scs\rm{{tfqmr}}}$&${\scs{{-}}}$&${\scs{{13}}}$&${\scs{{1850.3}}}$&${\scs{{3.5}}}$&${\scs{{524.7}}}$&${\scs{{3.5}}}$&${\scs{{3.5}}}$&${\scs{{795.1}}}$&${\scs{{2.3}}}$&${\scs{{4.3}}}$&${\scs{{1.1}}}$ \\
\hline
\end{tabular}
\end{center}
\end{table*}
\normalsize

\begin{table*}[p]
\begin{center}
\begin{tabular}[t]{c|c|c|c|ccc|c|}
\cline{2-8}
&${\scs\rm\bf{{cpus}}}$&${\scs\rm\bf{{Iteration}}}$&${\scs\rm\bf{{Product}}}$&\multicolumn{3}{|c|}{$\scs\rm\bf{{Tensor\:components}}$}&${\scs\rm\bf{{Memory}}}$ \\
&${\scs\rm\bf{{}}}$&${\scs\rm\bf{{}}}$&${\scs\rm\bf{{}}}$&${\scs\rm\bf{{Creation}}}$&${\scs\rm\bf{{DFT}}}$&${\scs\rm\bf{{Total}}}$&${\scs\rm\bf{{}}}$ \\
\cline{2-8}
\fcolorbox{black}{white}{${\scs\bf\color{black} BiCGSTAB}$}&${\scs\rm\bf{{}}}$&${\scs\rm\bf{{(s)}}}$&${\scs\rm\bf{{(s)}}}$&${\scs\rm\bf{{(s)}}}$&${\scs\rm\bf{{(s)}}}$&${\scs\rm\bf{{(s)}}}$&${\scs\rm\bf{{(GB)}}}$ \\
\hline\hline
${\scs\rm\bf{{ADDA}}}$&${\scs\rm\bf{{1}}}$&${\scs\rm\bf{{77.3}}}$&${\scs\rm\bf{{37.6}}}$&${\scs\rm\bf{{}}}$&${\scs\rm\bf{{}}}$&${\scs\rm\bf{{65.5}}}$&${\scs\rm\bf{{3.7}}}$ \\
${\scs\rm\bf{{openDDA\_S}}}$&${\scs\rm\bf{{1}}}$&${\scs\rm\bf{{68.3}}}$&${\scs\rm\bf{{32.7}}}$&${\scs\rm\bf{{1.8}}}$&${\scs\rm\bf{{11.6}}}$&${\scs\rm\bf{{13.4}}}$&${\scs\rm\bf{{3.1}}}$ \\
${\scs\rm\bf{{openDDA\_OMP}}}$&${\scs\rm\bf{{4}}}$&${\scs\rm\bf{{19.8}}}$&${\scs\rm\bf{{9.2}}}$&${\scs\rm\bf{{0.5}}}$&${\scs\rm\bf{{3.3}}}$&${\scs\rm\bf{{3.8}}}$&${\scs\rm\bf{{3.2}}}$ \\
\hline\hline
\\
\cline{2-8}
\fcolorbox{black}{white}{${\scs\bf\color{black} {\bf BICG}_{{\bf sym}}}$}&${\scs\rm\bf{{}}}$&${\scs\rm\bf{{(s)}}}$&${\scs\rm\bf{{(s)}}}$&${\scs\rm\bf{{(s)}}}$&${\scs\rm\bf{{(s)}}}$&${\scs\rm\bf{{(s)}}}$&${\scs\rm\bf{{(GB)}}}$ \\
\hline\hline
${\scs\rm\bf{{ADDA}}}$&${\scs\rm\bf{{1}}}$&${\scs\rm\bf{{40.2}}}$&${\scs\rm\bf{{37.6}}}$&${\scs\rm\bf{{}}}$&${\scs\rm\bf{{}}}$&${\scs\rm\bf{{65.3}}}$&${\scs\rm\bf{{3.1}}}$ \\
${\scs\rm\bf{{openDDA\_S}}}$&${\scs\rm\bf{{1}}}$&${\scs\rm\bf{{33.9}}}$&${\scs\rm\bf{{32.4}}}$&${\scs\rm\bf{{1.8}}}$&${\scs\rm\bf{{11.6}}}$&${\scs\rm\bf{{13.3}}}$&${\scs\rm\bf{{2.6}}}$ \\
${\scs\rm\bf{{openDDA\_OMP}}}$&${\scs\rm\bf{{4}}}$&${\scs\rm\bf{{10.5}}}$&${\scs\rm\bf{{9.6}}}$&${\scs\rm\bf{{0.5}}}$&${\scs\rm\bf{{3.3}}}$&${\scs\rm\bf{{3.8}}}$&${\scs\rm\bf{{2.6}}}$ \\
\hline\hline
\\
\cline{2-8}
\fcolorbox{black}{white}{${\scs\bf\color{black} {\bf QMR}_{{\bf sym}}}$}&${\scs\rm\bf{{}}}$&${\scs\rm\bf{{(s)}}}$&${\scs\rm\bf{{(s)}}}$&${\scs\rm\bf{{(s)}}}$&${\scs\rm\bf{{(s)}}}$&${\scs\rm\bf{{(s)}}}$&${\scs\rm\bf{{(GB)}}}$ \\
\hline\hline
${\scs\rm\bf{{ADDA}}}$&${\scs\rm\bf{{1}}}$&${\scs\rm\bf{{40.7}}}$&${\scs\rm\bf{{38.4}}}$&${\scs\rm\bf{{}}}$&${\scs\rm\bf{{}}}$&${\scs\rm\bf{{65.5}}}$&${\scs\rm\bf{{3.7}}}$ \\
${\scs\rm\bf{{openDDA\_S}}}$&${\scs\rm\bf{{1}}}$&${\scs\rm\bf{{35.1}}}$&${\scs\rm\bf{{33.1}}}$&${\scs\rm\bf{{1.8}}}$&${\scs\rm\bf{{11.6}}}$&${\scs\rm\bf{{13.4}}}$&${\scs\rm\bf{{3.1}}}$ \\
${\scs\rm\bf{{openDDA\_OMP}}}$&${\scs\rm\bf{{4}}}$&${\scs\rm\bf{{10.2}}}$&${\scs\rm\bf{{9.2}}}$&${\scs\rm\bf{{0.5}}}$&${\scs\rm\bf{{3.3}}}$&${\scs\rm\bf{{3.8}}}$&${\scs\rm\bf{{3.2}}}$ \\
\hline\hline
\end{tabular}
\caption{\label{tablenemoadda200} Comparative benchmark run on NEMO between ADDA 0.76 and the double-precision serial and OpenMP-based shared-memory variants of OpenDDA using ${\rm BiCGSTAB}$, ${\rm BICG}_{{\rm sym}}$ \& ${\rm QMR}_{{\rm sym}}$, for $K\!=\!J\!=\!P\!=\!200$, which produced a computational domain with $N\!=\!8\times\!10^{6}$ sites in total and a representation of the spherical target with $N_{d}\!=\!4,188,896$ occupied sites.}
\end{center}
\end{table*}

\begin{table*}[p]
\begin{center}
\begin{tabular}[t]{*{4}{c}}
\cline{2-4}
&\multicolumn{3}{c}{${\scs\bf SYSTEM\:SIZES\:FOR\:THE\:COMPARISON}$} \\
\cline{2-4}
${\scs\bf Nodes}$&${\scs\bf K=J=P}$&${\scs\bf N}$&${\scs\bf N_{d}}$ \\
\hline\hline
${\scs 4}$&${\scs 240}$&${\scs 13,824,000}$&${\scs 7,238,592}$ \\
${\scs 8}$&${\scs 310}$&${\scs 29,791,000}$&${\scs 15,599,512}$ \\
${\scs 16}$&${\scs 380}$&${\scs 54,872,000}$&${\scs 28,732,192}$ \\
${\scs 32}$&${\scs 480}$&${\scs 110,592,000}$&${\scs 57,904,928}$ \\
${\scs 64}$&${\scs 600}$&${\scs 216,000,000}$&${\scs 113,098,912}$ \\
\hline\hline
\end{tabular}
\caption{\label{tablesyssizewaltadda} System size, associated size of the computational domain and the number of occupied sites in the computational domain for the MPI-based comparative benchmark with ADDA across on 4, 8, 16, 32 \& 64 nodes on WALTON.}
\end{center}
\end{table*}

\begin{table*}[p]
\begin{center}
\begin{tabular}[t]{c|c|c|c|ccc|cc|}
\cline{2-9}
&${\scs\rm\bf{{cpus}}}$&${\scs\rm\bf{{Iteration}}}$&${\scs\rm\bf{{Product}}}$&\multicolumn{3}{|c|}{$\scs\rm\bf{{Tensor\:components}}$}&\multicolumn{2}{|c|}{$\scs\rm\bf{{Memory}}$} \\
&${\scs\rm\bf{{}}}$&${\scs\rm\bf{{}}}$&${\scs\rm\bf{{}}}$&${\scs\rm\bf{{Creation}}}$&${\scs\rm\bf{{DFT}}}$&${\scs\rm\bf{{Total}}}$&${\scs\rm\bf{{Total}}}$&${\scs\rm\bf{{Per\:node}}}$ \\
\cline{2-9}
\fcolorbox{black}{white}{${\scs\bf\color{black} BiCGSTAB}$}&${\scs\rm\bf{{}}}$&${\scs\rm\bf{{(s)}}}$&${\scs\rm\bf{{(s)}}}$&${\scs\rm\bf{{(s)}}}$&${\scs\rm\bf{{(s)}}}$&${\scs\rm\bf{{(s)}}}$&${\scs\rm\bf{{(GB)}}}$&${\scs\rm\bf{{(GB)}}}$ \\
\hline\hline
${\scs\rm\bf{{ADDA}}}$&${\scs\rm\bf{{4}}}$&${\scs\rm\bf{{64.8}}}$&${\scs\rm\bf{{31.2}}}$&${\scs\rm\bf{{}}}$&${\scs\rm\bf{{}}}$&${\scs\rm\bf{{41.7}}}$&${\scs\rm\bf{{7.6}}}$&${\scs\rm\bf{{2.2}}}$ \\
${\scs\rm\bf{{openDDA\_MPI}}}$&${\scs\rm\bf{{4}}}$&${\scs\rm\bf{{54.5}}}$&${\scs\rm\bf{{25.9}}}$&${\scs\rm\bf{{17.4}}}$&${\scs\rm\bf{{0.9}}}$&${\scs\rm\bf{{18.2}}}$&${\scs\rm\bf{{6.7}}}$&${\scs\rm\bf{{1.7}}}$ \\
\hline
${\scs\rm\bf{{ADDA}}}$&${\scs\rm\bf{{8}}}$&${\scs\rm\bf{{76.8}}}$&${\scs\rm\bf{{36.8}}}$&${\scs\rm\bf{{}}}$&${\scs\rm\bf{{}}}$&${\scs\rm\bf{{49.9}}}$&${\scs\rm\bf{{15.4}}}$&${\scs\rm\bf{{2.3}}}$ \\
${\scs\rm\bf{{openDDA\_MPI}}}$&${\scs\rm\bf{{8}}}$&${\scs\rm\bf{{63.9}}}$&${\scs\rm\bf{{30.6}}}$&${\scs\rm\bf{{24.4}}}$&${\scs\rm\bf{{0.9}}}$&${\scs\rm\bf{{25.3}}}$&${\scs\rm\bf{{14.7}}}$&${\scs\rm\bf{{1.8}}}$ \\
\hline
${\scs\rm\bf{{ADDA}}}$&${\scs\rm\bf{{16}}}$&${\scs\rm\bf{{79.9}}}$&${\scs\rm\bf{{36.8}}}$&${\scs\rm\bf{{}}}$&${\scs\rm\bf{{}}}$&${\scs\rm\bf{{45.2}}}$&${\scs\rm\bf{{27.6}}}$&${\scs\rm\bf{{2.1}}}$ \\
${\scs\rm\bf{{openDDA\_MPI}}}$&${\scs\rm\bf{{16}}}$&${\scs\rm\bf{{66.1}}}$&${\scs\rm\bf{{31.0}}}$&${\scs\rm\bf{{19.4}}}$&${\scs\rm\bf{{0.8}}}$&${\scs\rm\bf{{20.3}}}$&${\scs\rm\bf{{27.5}}}$&${\scs\rm\bf{{1.7}}}$ \\
\hline
${\scs\rm\bf{{ADDA}}}$&${\scs\rm\bf{{32}}}$&${\scs\rm\bf{{81.5}}}$&${\scs\rm\bf{{38.2}}}$&${\scs\rm\bf{{}}}$&${\scs\rm\bf{{}}}$&${\scs\rm\bf{{44.6}}}$&${\scs\rm\bf{{54.3}}}$&${\scs\rm\bf{{2.0}}}$ \\
${\scs\rm\bf{{openDDA\_MPI}}}$&${\scs\rm\bf{{32}}}$&${\scs\rm\bf{{71.6}}}$&${\scs\rm\bf{{34.1}}}$&${\scs\rm\bf{{23.3}}}$&${\scs\rm\bf{{0.8}}}$&${\scs\rm\bf{{24.1}}}$&${\scs\rm\bf{{54.4}}}$&${\scs\rm\bf{{1.7}}}$ \\
\hline
${\scs\rm\bf{{ADDA}}}$&${\scs\rm\bf{{64}}}$&${\scs\rm\bf{{98.1}}}$&${\scs\rm\bf{{46.9}}}$&${\scs\rm\bf{{}}}$&${\scs\rm\bf{{}}}$&${\scs\rm\bf{{62.9}}}$&${\scs\rm\bf{{116.7}}}$&${\scs\rm\bf{{2.2}}}$ \\
${\scs\rm\bf{{openDDA\_MPI}}}$&${\scs\rm\bf{{64}}}$&${\scs\rm\bf{{92.2}}}$&${\scs\rm\bf{{43.7}}}$&${\scs\rm\bf{{39.4}}}$&${\scs\rm\bf{{0.8}}}$&${\scs\rm\bf{{40.2}}}$&${\scs\rm\bf{{108.9}}}$&${\scs\rm\bf{{1.7}}}$ \\
\hline\hline
\\
\cline{2-9}
\fcolorbox{black}{white}{${\scs\bf\color{black} {\bf BICG}_{{\bf sym}}}$}&${\scs\rm\bf{{}}}$&${\scs\rm\bf{{(s)}}}$&${\scs\rm\bf{{(s)}}}$&${\scs\rm\bf{{(s)}}}$&${\scs\rm\bf{{(s)}}}$&${\scs\rm\bf{{(s)}}}$&${\scs\rm\bf{{(GB)}}}$&${\scs\rm\bf{{(GB)}}}$ \\
\hline\hline
${\scs\rm\bf{{ADDA}}}$&${\scs\rm\bf{{4}}}$&${\scs\rm\bf{{33.1}}}$&${\scs\rm\bf{{30.6}}}$&${\scs\rm\bf{{}}}$&${\scs\rm\bf{{}}}$&${\scs\rm\bf{{41.2}}}$&${\scs\rm\bf{{6.7}}}$&${\scs\rm\bf{{1.8}}}$ \\
${\scs\rm\bf{{openDDA\_MPI}}}$&${\scs\rm\bf{{4}}}$&${\scs\rm\bf{{27.3}}}$&${\scs\rm\bf{{25.9}}}$&${\scs\rm\bf{{17.3}}}$&${\scs\rm\bf{{0.9}}}$&${\scs\rm\bf{{18.2}}}$&${\scs\rm\bf{{5.7}}}$&${\scs\rm\bf{{1.4}}}$ \\
\hline
${\scs\rm\bf{{ADDA}}}$&${\scs\rm\bf{{8}}}$&${\scs\rm\bf{{37.6}}}$&${\scs\rm\bf{{35.5}}}$&${\scs\rm\bf{{}}}$&${\scs\rm\bf{{}}}$&${\scs\rm\bf{{49.2}}}$&${\scs\rm\bf{{13.3}}}$&${\scs\rm\bf{{1.9}}}$ \\
${\scs\rm\bf{{openDDA\_MPI}}}$&${\scs\rm\bf{{8}}}$&${\scs\rm\bf{{32.3}}}$&${\scs\rm\bf{{30.5}}}$&${\scs\rm\bf{{24.2}}}$&${\scs\rm\bf{{0.9}}}$&${\scs\rm\bf{{25.1}}}$&${\scs\rm\bf{{12.6}}}$&${\scs\rm\bf{{1.6}}}$ \\
\hline
${\scs\rm\bf{{ADDA}}}$&${\scs\rm\bf{{16}}}$&${\scs\rm\bf{{39.9}}}$&${\scs\rm\bf{{38.2}}}$&${\scs\rm\bf{{}}}$&${\scs\rm\bf{{}}}$&${\scs\rm\bf{{45.8}}}$&${\scs\rm\bf{{23.8}}}$&${\scs\rm\bf{{1.7}}}$ \\
${\scs\rm\bf{{openDDA\_MPI}}}$&${\scs\rm\bf{{16}}}$&${\scs\rm\bf{{30.1}}}$&${\scs\rm\bf{{30.4}}}$&${\scs\rm\bf{{19.9}}}$&${\scs\rm\bf{{0.8}}}$&${\scs\rm\bf{{20.8}}}$&${\scs\rm\bf{{23.6}}}$&${\scs\rm\bf{{1.5}}}$ \\
\hline
${\scs\rm\bf{{ADDA}}}$&${\scs\rm\bf{{32}}}$&${\scs\rm\bf{{40.6}}}$&${\scs\rm\bf{{37.9}}}$&${\scs\rm\bf{{}}}$&${\scs\rm\bf{{}}}$&${\scs\rm\bf{{45.7}}}$&${\scs\rm\bf{{45.6}}}$&${\scs\rm\bf{{1.7}}}$ \\
${\scs\rm\bf{{openDDA\_MPI}}}$&${\scs\rm\bf{{32}}}$&${\scs\rm\bf{{35.8}}}$&${\scs\rm\bf{{34.1}}}$&${\scs\rm\bf{{23.0}}}$&${\scs\rm\bf{{0.8}}}$&${\scs\rm\bf{{23.8}}}$&${\scs\rm\bf{{46.6}}}$&${\scs\rm\bf{{1.5}}}$ \\
\hline
${\scs\rm\bf{{ADDA}}}$&${\scs\rm\bf{{64}}}$&${\scs\rm\bf{{50.5}}}$&${\scs\rm\bf{{47.1}}}$&${\scs\rm\bf{{}}}$&${\scs\rm\bf{{}}}$&${\scs\rm\bf{{61.9}}}$&${\scs\rm\bf{{101.5}}}$&${\scs\rm\bf{{1.8}}}$ \\
${\scs\rm\bf{{openDDA\_MPI}}}$&${\scs\rm\bf{{64}}}$&${\scs\rm\bf{{45.8}}}$&${\scs\rm\bf{{43.2}}}$&${\scs\rm\bf{{38.4}}}$&${\scs\rm\bf{{0.8}}}$&${\scs\rm\bf{{39.2}}}$&${\scs\rm\bf{{93.7}}}$&${\scs\rm\bf{{1.5}}}$ \\
\hline\hline
\\
\cline{2-9}
\fcolorbox{black}{white}{${\scs\bf\color{black} {\bf QMR}_{{\bf sym}}}$}&${\scs\rm\bf{{}}}$&${\scs\rm\bf{{(s)}}}$&${\scs\rm\bf{{(s)}}}$&${\scs\rm\bf{{(s)}}}$&${\scs\rm\bf{{(s)}}}$&${\scs\rm\bf{{(s)}}}$&${\scs\rm\bf{{(GB)}}}$&${\scs\rm\bf{{(GB)}}}$ \\
\hline\hline
${\scs\rm\bf{{ADDA}}}$&${\scs\rm\bf{{4}}}$&${\scs\rm\bf{{33.2}}}$&${\scs\rm\bf{{32.5}}}$&${\scs\rm\bf{{}}}$&${\scs\rm\bf{{}}}$&${\scs\rm\bf{{41.8}}}$&${\scs\rm\bf{{7.6}}}$&${\scs\rm\bf{{2.2}}}$ \\
${\scs\rm\bf{{openDDA\_MPI}}}$&${\scs\rm\bf{{4}}}$&${\scs\rm\bf{{27.7}}}$&${\scs\rm\bf{{26.0}}}$&${\scs\rm\bf{{17.6}}}$&${\scs\rm\bf{{0.9}}}$&${\scs\rm\bf{{18.4}}}$&${\scs\rm\bf{{6.7}}}$&${\scs\rm\bf{{1.7}}}$ \\
\hline
${\scs\rm\bf{{ADDA}}}$&${\scs\rm\bf{{8}}}$&${\scs\rm\bf{{39.2}}}$&${\scs\rm\bf{{35.9}}}$&${\scs\rm\bf{{}}}$&${\scs\rm\bf{{}}}$&${\scs\rm\bf{{48.1}}}$&${\scs\rm\bf{{15.4}}}$&${\scs\rm\bf{{2.3}}}$ \\
${\scs\rm\bf{{openDDA\_MPI}}}$&${\scs\rm\bf{{8}}}$&${\scs\rm\bf{{32.8}}}$&${\scs\rm\bf{{30.2}}}$&${\scs\rm\bf{{24.5}}}$&${\scs\rm\bf{{0.9}}}$&${\scs\rm\bf{{25.3}}}$&${\scs\rm\bf{{14.7}}}$&${\scs\rm\bf{{1.8}}}$ \\
\hline
${\scs\rm\bf{{ADDA}}}$&${\scs\rm\bf{{16}}}$&${\scs\rm\bf{{41.0}}}$&${\scs\rm\bf{{37.4}}}$&${\scs\rm\bf{{}}}$&${\scs\rm\bf{{}}}$&${\scs\rm\bf{{46.3}}}$&${\scs\rm\bf{{27.6}}}$&${\scs\rm\bf{{2.1}}}$ \\
${\scs\rm\bf{{openDDA\_MPI}}}$&${\scs\rm\bf{{16}}}$&${\scs\rm\bf{{33.3}}}$&${\scs\rm\bf{{31.2}}}$&${\scs\rm\bf{{20.1}}}$&${\scs\rm\bf{{0.8}}}$&${\scs\rm\bf{{20.9}}}$&${\scs\rm\bf{{27.5}}}$&${\scs\rm\bf{{1.7}}}$ \\
\hline
${\scs\rm\bf{{ADDA}}}$&${\scs\rm\bf{{32}}}$&${\scs\rm\bf{{40.3}}}$&${\scs\rm\bf{{38.1}}}$&${\scs\rm\bf{{}}}$&${\scs\rm\bf{{}}}$&${\scs\rm\bf{{46.5}}}$&${\scs\rm\bf{{54.3}}}$&${\scs\rm\bf{{2.0}}}$ \\
${\scs\rm\bf{{openDDA\_MPI}}}$&${\scs\rm\bf{{32}}}$&${\scs\rm\bf{{37.6}}}$&${\scs\rm\bf{{35.3}}}$&${\scs\rm\bf{{22.7}}}$&${\scs\rm\bf{{0.8}}}$&${\scs\rm\bf{{23.5}}}$&${\scs\rm\bf{{54.4}}}$&${\scs\rm\bf{{1.7}}}$ \\
\hline
${\scs\rm\bf{{ADDA}}}$&${\scs\rm\bf{{64}}}$&${\scs\rm\bf{{50.8}}}$&${\scs\rm\bf{{47.5}}}$&${\scs\rm\bf{{}}}$&${\scs\rm\bf{{}}}$&${\scs\rm\bf{{61.4}}}$&${\scs\rm\bf{{116.7}}}$&${\scs\rm\bf{{2.2}}}$ \\
${\scs\rm\bf{{openDDA\_MPI}}}$&${\scs\rm\bf{{64}}}$&${\scs\rm\bf{{46.9}}}$&${\scs\rm\bf{{43.8}}}$&${\scs\rm\bf{{38.2}}}$&${\scs\rm\bf{{0.8}}}$&${\scs\rm\bf{{39.0}}}$&${\scs\rm\bf{{108.9}}}$&${\scs\rm\bf{{1.7}}}$ \\
\hline\hline
\end{tabular}
\caption{\label{tablewaltonadda} Comparative benchmark run on WALTON between ADDA 0.76 and the double-precision MPI-based distributed-memory variant of OpenDDA using ${\rm BiCGSTAB}$, ${\rm BICG}_{{\rm sym}}$ \& ${\rm QMR}_{{\rm sym}}$, across 4, 8, 16, 32 \& 64 nodes.}
\end{center}
\end{table*}

\begin{table*}[p]
\begin{center}
\begin{tabular}[t]{cccc}
\hline
\multicolumn{1}{c}{${\scs\bf DDSCAT}$}&\multicolumn{1}{c}{${\scs\bf OpenDDA\:(Serial)}$}&\multicolumn{1}{c}{${\scs\bf OpenDDA\:(OpenMP[4])}$}&\multicolumn{1}{c}{${\scs\bf OpenDDA\:(MPI[4])}$} \\
\hline\hline
${\scs Q_{ext}=1.264}$&${\scs Q_{ext}=1.265}$&${\scs Q_{ext}=1.265}$&${\scs Q_{ext}=1.265}$ \\
${\scs Q_{abs}=0.911}$&${\scs Q_{abs}=0.911}$&${\scs Q_{abs}=0.911}$&${\scs Q_{abs}=0.911}$ \\
${\scs Q_{sca}=0.353}$&${\scs Q_{sca}=0.354}$&${\scs Q_{sca}=0.354}$&${\scs Q_{sca}=0.354}$ \\
\hline\hline
\end{tabular}
\caption{\label{tableopenddavalid} Results for the operational verification of the three architectural variants of the new OpenDDA framework. Parameters: {\sc CUBE}, $K\!=\!J\!=P\!=\!100\Rightarrow N=N_{d}=1\!\times\!10^{6}$, normal incidence on one face of the cube, incident wavelength $\lambda=3.175\:\mu{\rm m}$, incident polarisation ${\rm {\hat e}_{0}}={\rm {\hat x}}$, effective radius $a=0.5\:\mu{\rm m}$, and complex refractive index $m=1.63631+0.372i$.}
\end{center}
\end{table*}

\begin{figure}[p]
\begin{center}
\includegraphics[scale=0.5]{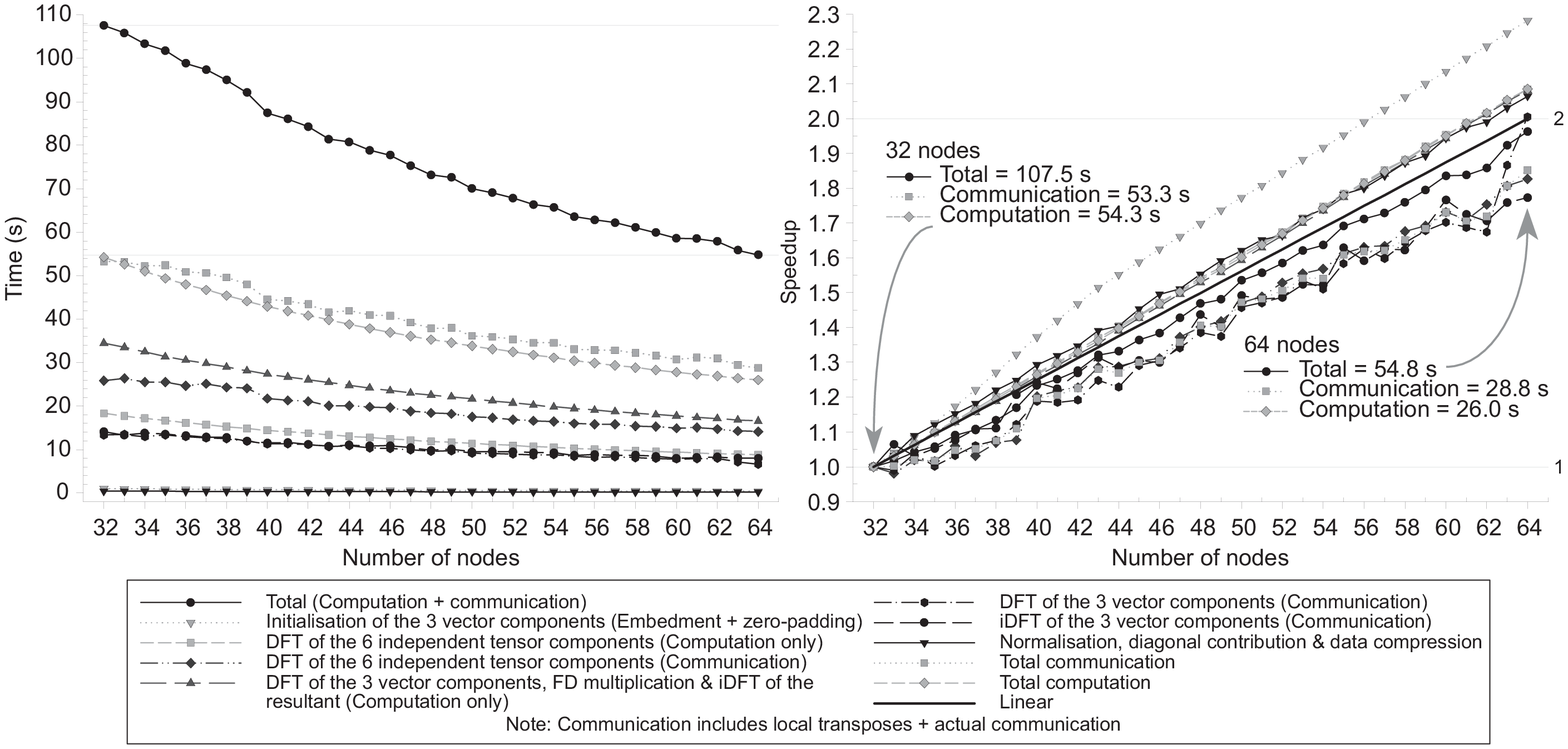}
\caption{\label{fig5}MPI scalability and timing run on WALTON with associated speedups for a spherical target with $K\!=\!J\!=\!P\!=\!600$ and $N_{d}\!=\!113,089,912$ occupied sites.}
\end{center}
\end{figure}

\begin{figure}[p]
\begin{center}
\includegraphics[scale=0.3]{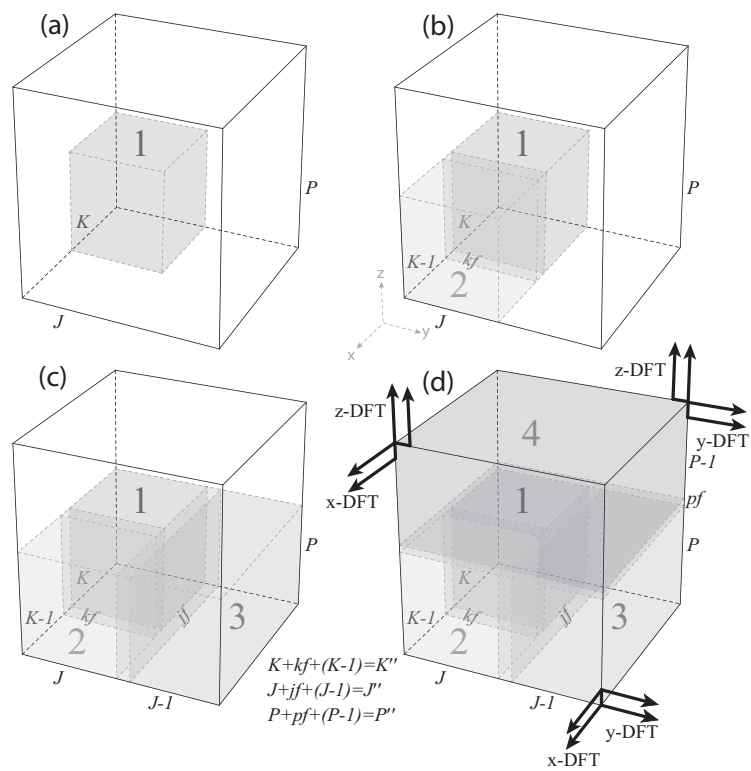}
\vspace{-1em}
\caption{\label{fig6} Outlines the T-to-c conversion of the six tensor-component arrays, the zero-padding of the three vector-component arrays and subsequent DFTs \& iDFT for the original algorithm.}
\end{center}
\end{figure}

\newpage
\begin{center}{\bf\Large Biographies}\end{center}
James Mc Donald received his B.Sc. (1st Class Honours) in Experimental Physics from the National University of Galway (NUI, Galway) in 2002. Subsequently, he worked with the Semiconductor Physics Group in the School of Physics and Astronomy at the University of St Andrews. He returned to Galway to undertake a Ph.D. under the supervision of Dr. Golden and Prof. Jennings at NUI, Galway, which he completed in late 2007. His research interests include high-performance computing and astrophysics (light scattering by non-spherical particles). James Mc Donald gained his Ph.D. in Computational Physics from the National University of Ireland in 2008.\\\vspace{1em}

Aaron Golden is College Lecturer in the Department of Information Technology at the National University of Ireland, Galway, and a member of the University's National Centre for Biomedical Engineering Science, and the Centre for Astronomy. He read Natural Sciences at Trinity College Dublin gaining a BA (Mod. Experimental Physics) in 1991, an M.Sc. in Computational Science from the Queen's University of Belfast (1993), and the Ph.D. degree in Astrophysics from the National University of Ireland in 1999. His research interests span topics in astrophysics (magnetospheric emission processes in pulsars and ultracool dwarf stars), computational biology (functional annotation of genomes using artificial intelligence techniques) and high performance computation. This is reflected in the range of journals he has had his work published, from the Astrophysical Journal to Nucleic Acids Research, from Bioinformatics, to Science with 26 peer reviewed published articles and numerous more at conference level. Dr. Golden holds two active Research Frontier Programme Awards from Science Foundation Ireland in Physics and Computer Science, and has supervised 9 Ph.D. students in the 8 years since his appointment at NUI Galway.\\\vspace{1em}

S. Gerard Jennings graduated from the National University of Ireland, Galway (NUI, Galway) in 1965 in Experimental Physics with a B.Sc (1st. Class honours), followed by a M.Sc. in 1967. He obtained a Ph.D. in Atmospheric Physics from the University of Manchester in 1971. In 1980, he was appointed Statutory Lecturer in the Department of Experimental Physics, NUI, Galway. He was awarded a D.Sc. from the University of Manchester, 1987. Professor of Experimental Physics at NUI, Galway (1990-present). He was awarded Doctor Honoris Causa by the University of G\"oteborg in 2000. He was Head of Department of Experimental Physics (1992-1995; 2002-2005) at NUI, Galway, and is currently the director of the Environmental Change Institute (ECI) at NUI, Galway. He is the director of the Atmospheric Research Group, leader of the Climate Change Cluster within the ECI, and management (Secretary) of the Mace Head Atmospheric Research Station, near Carna, Co. Galway. He is Chair of the UK Natural Environmental Research Council (NERC) Polluted Troposphere Scientific Steering Committee (2002-present), a member of the Scientific Advisory Group of the World Meteorological Organisation, Global Atmosphere Watch, (GAW) Aerosol (1997-present), a member of IGAC’s Coordinating Committee for Aerosol Characterisation and Process Studies Activities, Associate Editor of Journal of Geophysical Research (2002-2005), and a member of the Scientific Steering Committee of EUROTRAC (1992-1999). His research interests include atmospheric physics including aerosol physics, climate change, effects of aerosols on climate, atmospheric field studies at the Mace Head Atmospheric Research Station, environmental pollution, laser propagation in aerosol and cloud media and satellite remote sensing. He has 91 peer refereed journal publications, 1 book, 6 book chapters and 100 conference papers. He has supervised 10 Ph.D. and 14 research M.Sc. students.
\end{document}